%% file: soapcnn.tex
\documentclass[aps,prd,twocolumn,superscriptaddress,groupedaddress]{revtex4}
\usepackage{graphicx}  
\usepackage{dcolumn}   
\usepackage{bm}        
\usepackage{amssymb}   
\usepackage{amsmath}
\usepackage{color}
\usepackage{acronym}
\usepackage{algcompatible}
\usepackage{newfloat}
\usepackage{cancel}
\usepackage{tikz}
\usepackage{hyperref}
\usetikzlibrary{shadows,arrows,arrows.meta,positioning,backgrounds,fit,chains,scopes}
\usepackage{caption}
\usepackage[caption=false]{subfig}
\usepackage{changepage}
\graphicspath{{./figures/}}
\usepackage[T1]{fontenc}

\begin{document}

\title{A robust machine learning algorithm to search for continuous gravitational waves.}

\author{Joe Bayley}
\author{Chris Messenger}
\author{Graham Woan}
\affiliation{SUPA, University of Glasgow, Glasgow G12 8QQ, United Kingdom.}


\begin{abstract}
Many continuous gravitational wave searches are affected by instrumental
spectral lines that could be confused with a continuous astrophysical
signal. Several techniques have been developed to limit the effect of these
lines by penalising signals that appear in only a single detector. We have
developed a general method, using a convolutional neural network, to reduce
the impact of instrumental artefacts on searches that use the SOAP
algorithm \cite{bayley2019SOAPGeneralised}. The method can identify features in corresponding
frequency bands of each detector and classify these bands as containing a
signal, an instrumental line, or noise. We tested the method against four
different data-sets: Gaussian noise with time gaps, data from the final run
of Initial LIGO (S6) with signals added,  the reference S6 mock data
challenge data set \cite{walsh2016ComparisonMethods} and signals injected into data from the
second advanced LIGO observing run (O2). Using the S6 mock data challenge
data set and at a 1\% false alarm probability we showed that at 95\% efficiency a
fully-automated SOAP search has a sensitivity corresponding to a coherent
signal-to-noise ratio of 110, equivalent to a sensitivity depth of $10 \;
\rm{Hz}^{-1/2}$, making this automated search competitive with other
searches requiring significantly more computing resources and human
intervention.

\end{abstract}

\maketitle

\acrodef{GW}[GW]{gravitational-wave}
\acrodef{CW}[CW]{continuous gravitational wave}
\acrodef{NS}[NS]{neutron star}
\acrodef{EM}[EM]{electromagnetic}
\acrodef{SNR}[SNR]{signal-to-noise-ratio}
\acrodef{LIGO}[LIGO]{Laser Interferometer Gravitational-wave Observatory}
\acrodef{SFT}[SFT]{short Fourier transform}
\acrodef{FFT}[FFT]{fast Fourier transform}
\acrodef{UCD}[UCD]{up, centre or down}
\acrodef{MDC}[MDC]{mock data challenge}
\acrodef{PSD}[PSD]{power spectral density}
\acrodef{ROC}[ROC]{receiver operating characteristic}
\acrodef{RMS}[RMS]{root median square}
\acrodef{MCMC}[MCMC]{Markov-Chain Monte Carlo}
\acrodef{CNN}[CNN]{convolutional neural network}
\acrodef{CBC}[CBC]{compact binary coalescence}
\acrodef{RELU}[RELU]{rectified linear unit}
\acrodef{vitstat}[vitstat]{Viterbi statistic}
\acrodef{vitmap}[vitmap]{Viterbi map}
\acrodef{CPU}[CPU]{central processing unit}
\acrodef{GPU}[GPU]{graphics processing unit}

\section{\label{intro:1}Introduction}
%

%
Gravitational-wave detectors such as the
\ac{LIGO}~\cite{abbott2009LIGOLaser,aasi2015AdvancedLIGO} and Virgo
\cite{acernese2015AdvancedVirgo,acernese2008StatusVirgo} are sensitive to
signals from many types of astrophysical sources. One type, \acp{CBC}, has
already been observed in quantity
~\cite{abbott2017GW170817Observation,abbott2017GW170814ThreeDetector,abbott2016ObservationGravitational},
however, other promising source types, including  sources of \acp{CW}, remain
undetected. \acp{CW} are well-modelled quasi-sinusoidal signals with a
duration much longer than observing times of detectors. The sources of these
signals are thought to be rapidly rotating neutron stars, which will emit
\acp{GW} if there is some asymmetry around the rotation axis
\cite{prix2009GravitationalWaves}. The signals will have small amplitudes
compared to \acp{CBC}, and only detectable with sensitive algorithms and
observing times of months or years. These search algorithms are generally
classed as `targeted', `directed', or `all-sky' searches, dependent on how
much is known a priori about the source from \ac{EM} observations.

%
Targeted searches can be performed on sources with known sky position and
spin evolution.  If only the sky position is known one can perform a directed
search, and if nothing is known one is forced to perform an all-sky search
covering sky position and source rotational frequency (and usually frequency
derivative). The most sensitive of these are targeted searches which can
employ variants on coherent matched
filtering~\cite{dupuis2005BayesianEstimation,schutz1998DataAnalysis}. These
use template waveforms which are generated using the information already
known about the source, then correlated this with the data. Directed and
all-sky searches have a much broader parameter space to search, therefore,
many templates are needed to sufficiently cover the parameter space. Coherent
matched filter methods have a high computing burden in broader parameter
space searches and become unfeasible. This led to the development of
semi-coherent searches, in which the data is divided into segments that are
analysed separately and their results combined
incoherently~\cite{abbott2019AllskySearch,creighton2000SearchingPeriodic}.
Semi-coherent searches are generally tuned to deliver maximum sensitivity for
a given computing time.
An overview of current \ac{CW} searches can be found in \cite{riles2017December19, sieniawska2019ContinuousGravitational}.

%
The analysis presented here applies to an existing semi-coherent search
algorithm called SOAP~\cite{bayley2019SOAPGeneralised, bayley2020Soapcw}. This is a fast and
largely modelled-independent search which uses a Viterbi-like algorithm to
find continuous tracks of excess power in time-frequency spectrograms. When
applied to multiple detectors using a line-aware statistic, SOAP looks for
frequency bins which have consistent high power in each detector. This means
that, at a given frequency and a given time, SOAP will penalise signals which
are not seen consistently in the detector network. The algorithmic details
are summarised in Sec.~\ref{soap}.

%

The sensitivity of SOAP, and many other \ac{GW} searches, is limited by noise
artefacts known as `instrumental lines' which have been investigated in \cite{covas2018IdentificationMitigation} for advanced \ac{LIGO}.  This generic term covers a range of
artefacts, including long-duration, fixed-frequency or wandering lines to
fixed-frequency transients. There have been many techniques to mitigate the effect of these lines on searches including \cite{keitel2014SearchContinuousa,leaci2015MethodsFilter,zhu2017NewVeto}. For the SOAP search, there are certain types of instrumental line that
are difficult to distinguish from an astrophysical signal even with the
development of a `line aware' statistic~\cite{bayley2019SOAPGeneralised}.
Currently these require one to manually examine the problematic sub-bands to
determine whether they are contaminated. This process is slow and requires
significant human input and judgement, and for full-band, long-duration
searches it would become impractical.

%

We have therefore automated this process by employing \acp{CNN}. These have
been used extensively in image classification problems, and we explain their
use in more detail in Sec.~\ref{cnn}. \acp{CNN} have already been shown to
detect gravitational wave signals from \acp{CBC}
in~\cite{gabbard2018MatchingMatched,george2018DeepLearning,gebhard2019ConvolutionalNeural}, have been used in searches for burst signals \cite{chan2019DetectionClassification, iess2020CoreCollapseSupernova, astone2018NewMethodb} and various deep learning techniques have been used in searching for \ac{CW}
signals in~\cite{dreissigacker2019DeeplearningContinuous, astone2018NewMethodb, miller2019HowEffectivea, morawski2020ConvolutionalNeural}. 
An overview of machine learning techniques used in \ac{GW} science can be found in \cite{cuoco2020EnhancingGravitationalWave}.

%
In Sec.\ref{soap} we will summarise how the SOAP algorithm works. In
Sec.~\ref{cnn} we explain how \acp{CNN} operate, and we show how we generate
data to train the \ac{CNN} in Sec.~\ref{data}. We describe the entire search,
from raw data to results, in Sec.~\ref{pipeline} and finally in
Sec.~\ref{results} we show our results from real searches, comparing them to
corresponding results using other techniques.

\section{\label{soap} Soap}

%
SOAP~\cite{bayley2019SOAPGeneralised} is a search algorithm for un-modeled
long-duration signals based on the Viterbi
algorithm~\cite{viterbi1967ErrorBounds}. In its most simple form SOAP
analyses a spectrogram to find the continuous time-frequency track which
gives the highest sum of \ac{FFT} power. If a signal is present this is the
track which is most likely to correspond to that signal.
In~\cite{bayley2019SOAPGeneralised} the algorithm was expanded to include
multiple detectors as well as a statistic to penalise artefacts in the data
from instrumental lines.

%
Fig.~\ref{soap:viterbiplot} shows an example of a time-frequency spectrogram
and the corresponding outputs from SOAP; the three main output components are
the frequency track, the Viterbi statistic and the Viterbi map, described
below:

\begin{figure*}
\includegraphics[scale=0.48]{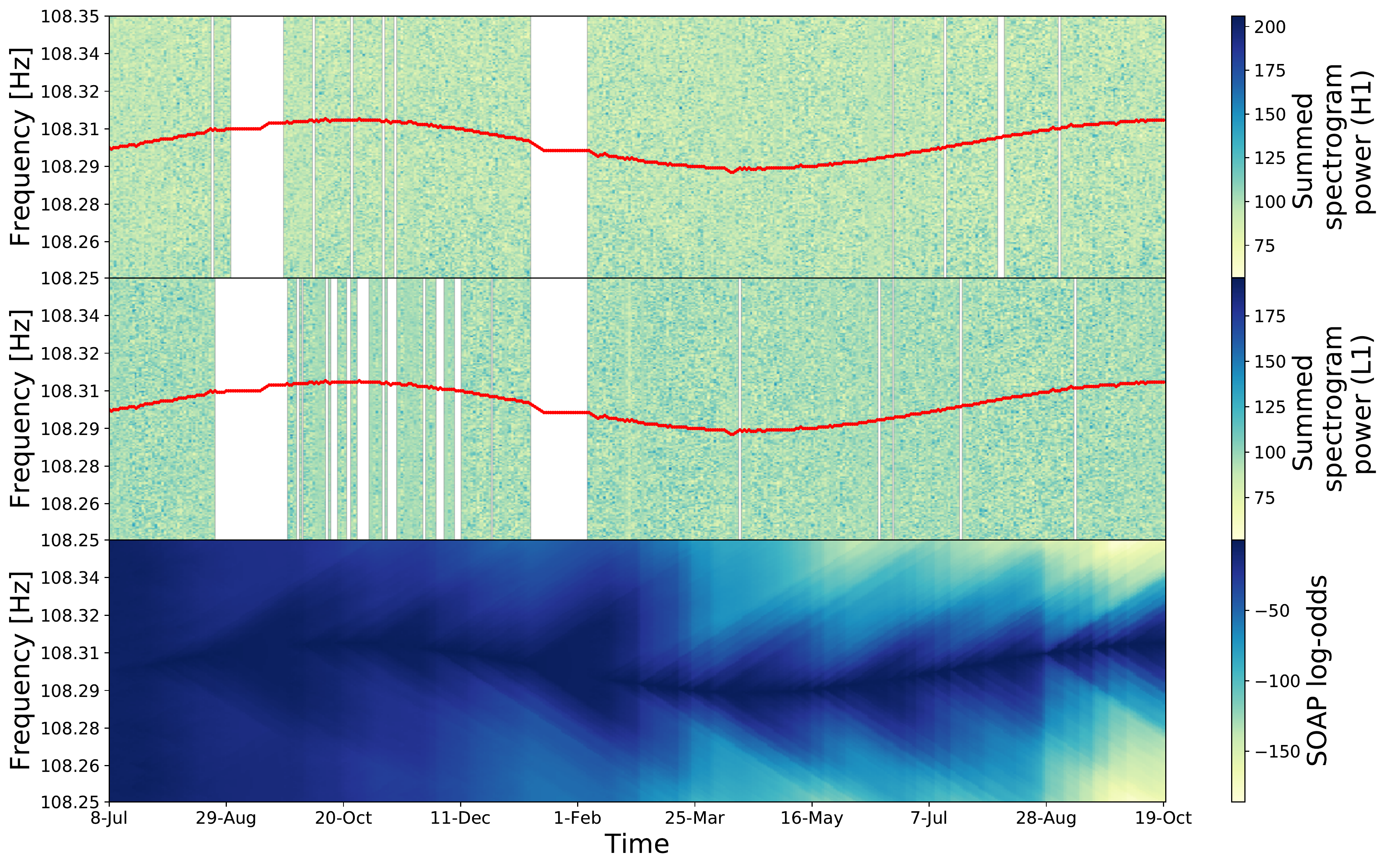}
\caption{\label{soap:viterbiplot} An example of a SOAP search. The top two
panels show time-frequency spectrograms, pre-processed according to
Sec.~\ref{pipeline} and representing a 0.1\,Hz-wide frequency band from the
\acp{LIGO} S6 observing run.  The data includes a simulated \ac{CW} signal.
The white areas in the spectrograms are gaps in data when the corresponding
detector was not operating. The optimal track found by SOAP is overlaid in
both cases. The bottom panel shows the normalised Viterbi map with the pixel
intensity showing the log-probability that the track falls in a particular
frequency bin as a function of time.}
\end{figure*}

%
\begin{description}
\item [Viterbi track] The Viterbi track is the most probable track through
    time-frequency data for given a choice of statistic (i.e. summed \ac{SFT} power).
\item [Viterbi statistic] The Viterbi statistic is the sum of the
    individual statistics along the Viterbi track. In the analysis that
    follows we use the `line-aware' Viterbi statistic. This is the sum of
    the log-odds ratios, $p_{\rm signal}/(p_{\rm line} + p_{\rm noise})$
    along the track \cite{bayley2019SOAPGeneralised}.
\item [Viterbi map] The Viterbi map shows the value of the Viterbi
    statistic for every time-frequency bin in the spectrogram,
    corresponding to the log-probability that the track passes through each
    time-frequency bin. Each time slice in the map is normalised
    individually, i.e., each vertical slice is adjusted so that the sum of
    their exponentiated values is unity. Each pixel in the image can
    therefore be interpreted as a value related to the log-probability that
    the signal is in that frequency bin at that time.

\end{description}

%
In~\cite{bayley2019SOAPGeneralised} we used the Viterbi `line-aware'
statistic (described above) to determine whether the signal had an
astrophysical origin. This statistic reduces the effect of instrumental lines
on the analysis, but certain types of line are not picked-up by it. For
example, the statistic is affected by broad, wandering, common lines as they
offer high power tracks in both detectors. To reduce the effect of these
instrumental lines in ~\cite{bayley2019SOAPGeneralised}, we examined the
spectrograms and Viterbi maps of individual bands by eye, as in
Fig.~\ref{soap:viterbiplot}. Bands which appeared to be contaminated were
then manually removed from the search.

%
In this paper we show that we can exploit additional information in the
spectrograms and Viterbi map, in combination with the Viterbi statistic, to
perform the process of removing contaminated bands automatically . The tool
which we use to classify this extra information is a convolutional neural
network.

\section{\label{cnn}Convolutional neural networks}

%
\acp{CNN} are a type of deep neural network which are primarily used in image
processing and recognition
\cite{lecun2015DeepLearning,lecun1998GradientbasedLearning,waibel1989PhonemeRecognition,krizhevsky2012ImageNetClassificationa}.
A \ac{CNN} is designed to take in data, identify different features within
that data and classify what those features or combinations of those features
mean. In the context of this work the input data is a time-frequency
spectrogram which may contain a (simulated) \ac{CW} signal. The output is
then a single number which represents the confidence that a signal is
present. Values closer to 1 represent the presence of a signal and closer to
0 represent its absence. A \ac{CNN} can learn how to identify features by
training on many labelled examples of the input data where the output is
known. For example, an input spectrogram with a \ac{CW} signal would have a
label of 1. Given the set of training examples, the many parameters of the
\ac{CNN} can be updated such that it gives the best result for any new
spectrogram image. The many parameters of the \ac{CNN} relate to the key
building block of neural networks: the neuron.

\subsection{\label{cnn:neurons} Neurons}
\begin{figure}
\includegraphics[width=\columnwidth]{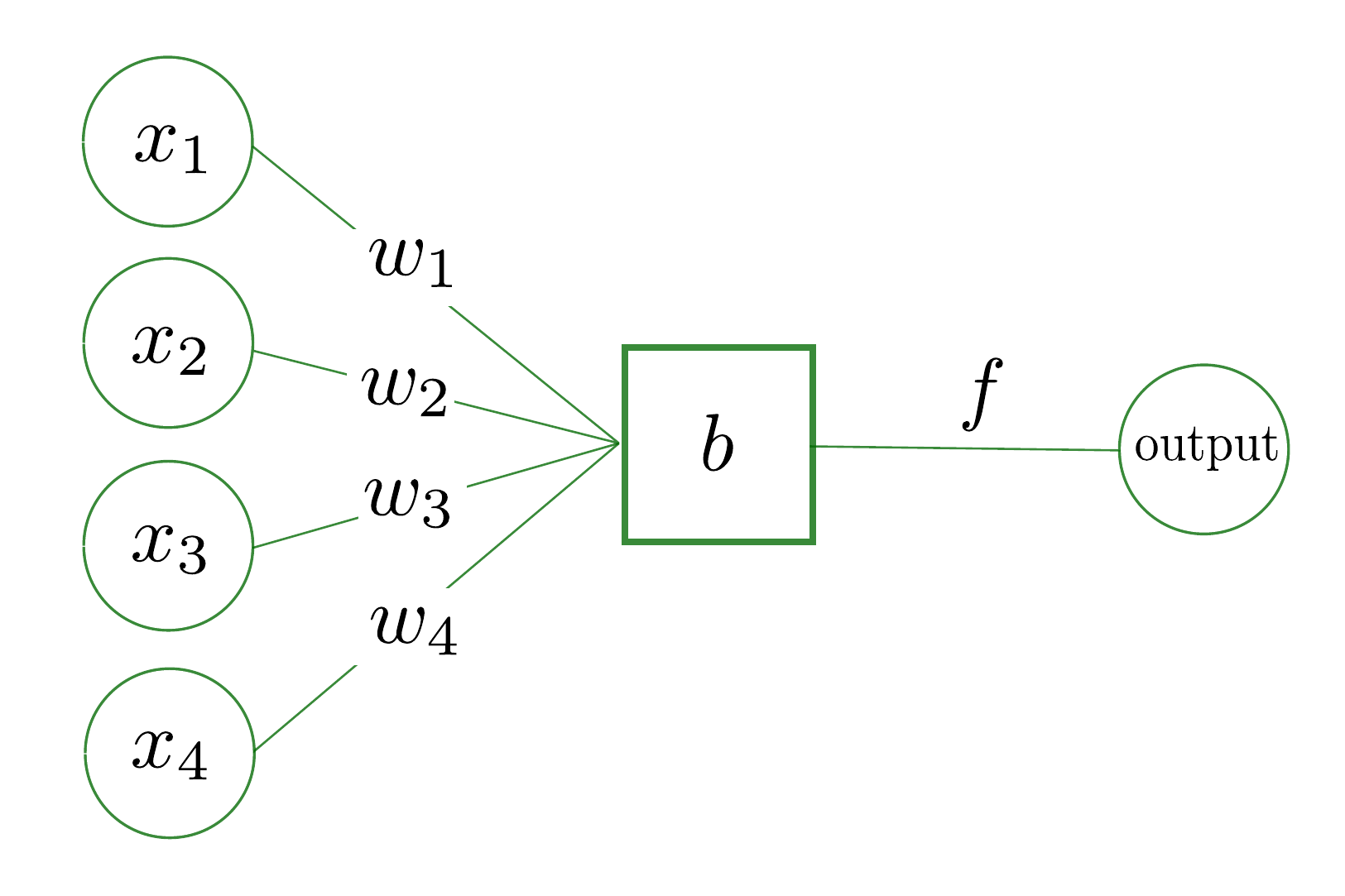}
\caption{\label{cnn:singleneuron} Eq.~\ref{machine:nn:neuron:equation} can be
visualized as above. Here the inputs $x_i$ are multiplied by the corresponding
weights $w_i$, the sum of these and the bias $b$ are then passed through an activation function $f$ to the output.
This example has four inputs but there can be any number.
}
\end{figure}

%
%
A neuron converts any number of inputs into a single output value and can
perform three operations which are applied to $N$ inputs ${ \bm x}$:
multiplying each input by a `weight' $w$, adding a `bias' $b$ and passing
them through, and applying a non-linear `activation function' $f$.
Fig.~\ref{cnn:singleneuron} shows this basic operation, where there is one
weight for every input and a single bias for each neuron. The output $O$ is
therefore related to the inputs by
\begin{equation}
    O = f\left(b + \sum_{i=1}^{N} w_i x_i  \right).
    \label{machine:nn:neuron:equation}
\end{equation}
The weights and bias are the parameters which the neural network `learns'
during training and we consider this further in Sec.~\ref{cnn:training}. The
activation function is there to explicitly impose a non-linearity to the
calculation.

\subsection{\label{cnn:nn}Neural Networks}

To create a neural network, many of these neurons are connected together into
`layers'. Layers comprise a set of neurons, each of which takes the same data
as input but which applies a different sets of weights and biases. The output
of each neuron then acts as the input to another set of neurons or another
layer. Neural networks combine many of these layers together to learn
abstract representations of the data. For classification, this abstract
representation is distilled down to a simple output. Neural networks can be
made from many different types of layers. We described a `fully-connected'
layer above, in which each neuron in the layer takes in all the data points
or the previous layer's outputs. However for certain types of problem, such
as identifying features in images, another type of layer called a
convolutional layer is better suited.

\subsection{\label{cnn:convolutional}Convolutional layers}

\begin{figure}
\includegraphics[width=\columnwidth]{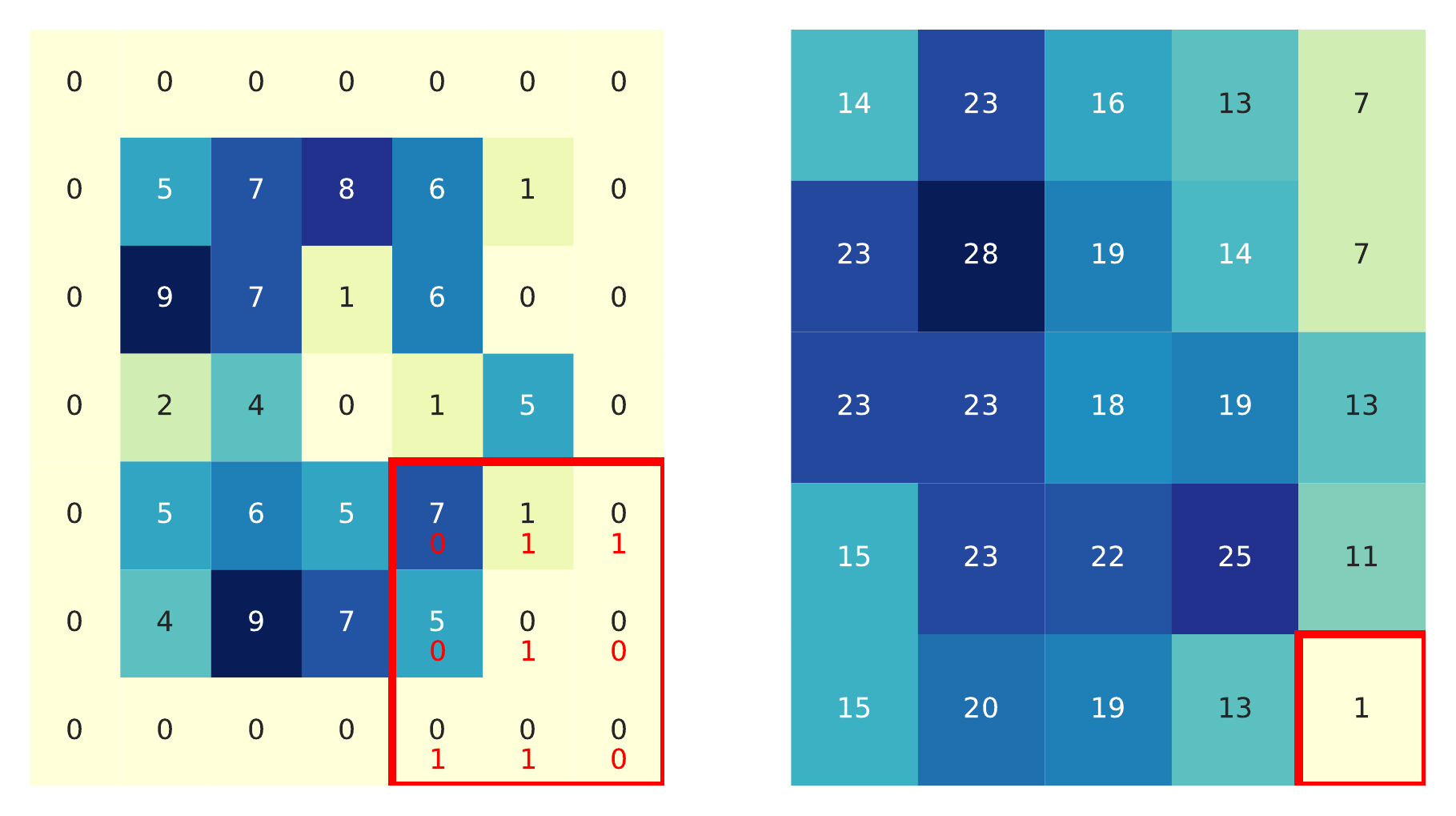}
\caption{\label{cnn:convlayer} The convolutional layers convolve a filter
with the input image and output a convolved image the same size as the input
to pass to the next layer. Here we show a simple $5\times5$ image with a
$3\times3$ filter, the input is padded with zeros such that the output is the
same size. When the network is trained, the values within the filter (the red
values below the inputs) are updated. }
\end{figure}

%
Convolutional layers are an adaptation of the fully-connected layers
described above in Sec.~\ref{cnn:nn}, where the input data is generally image pixels. For this type of layer there is not a
separate weight for each input data point (pixel). Rather, there is a fixed number of
weights defined by a `filter' size. This filter is convolved with the input
image such that the output of the
layer is a filtered image. This operation is shown in
Fig.~\ref{cnn:convlayer}. The convolutional layers equivalent to
Eq.~\ref{machine:nn:neuron:equation} is,
\begin{equation}
\label{cnn:conv:equation}
    O_{ij} = f\left(b +  \sum_{m} \sum_{n} F_{mn}x_{i-m, j-n}\right) ,
\end{equation}
where $O$ is the output image, $b$ is the bias,
$x$ is the input image, $F$ is the convolutional filter and $f$ is the
activation function. The indices $m$ and $n$ iterate over the filter rows and columns and the indices $i$ and $j$ iterate over the input image rows and columns. The convolutional layer can learn to identify features
within an image by changing the weights and bias of a filter. A
convolutional layer can apply a number of these filters as defined by the
user. If the layer has 10 filters then the output is 10 filtered images. Each
of these filters can then be trained to identify different features within
the input image \cite{lecun1998GradientbasedLearning,krizhevsky2012ImageNetClassificationa}.

%
The output of a convolutional layer comprises a number of filtered images, so
potentially there is a lot of data to feed to the next layer. A method called
max-pooling can be used to reduce the size of the output whilst retaining the
important information within the images. A max-pooling layer
splits the image into blocks of fixed size and takes the maximum
pixel value in each block as the output. So if the size of the
max-pooling block is $2\times2$, the output image will have $1/4$  the number
of pixels of the input.

%

\subsection{\label{cnn:training}Training}
%
Once the structure of the network is decided, the network needs to be trained
by adjusting the weights and biases to give the desired performance. To
achieve this the networks classify the input images (the spectrograms and Viterbi maps)
using a single output neuron with a sigmoid activation function which restricts the output between 0 and 1. The \ac{CNN} is trained using a supervised learning
process in which the class of each input example is known. We assign a label
of 1 when the input is a time-frequency spectrogram which includes a
simulated \ac{CW} signal and 0 when there is no signal.

The performance of the network can be improved by increasing the number of
input examples which is sees during training.  This  helps it learn the
underlying features within the data and prevents it from over-fitting to
specific examples.

%
Each of the training examples is then propagated though the network to its
corresponding single output value, which lies between 0 and 1. This output is
then compared to the label of the input data using a loss function. For our
two-class network the loss function, $L$, is the binary cross-entropy \cite{goodfellow2016DeepLearning}
\begin{equation}\label{cnn:loss}
L = -y\log{(p)} + (1-y)\log{(1-p)},
\end{equation}
where $p$ is the network's output, which has any value in the range $[0,1]$ and $y$ is the true
output which has the binary label 0 or 1. The loss function is minimised when
the output matches the truth. Its current value is used to train the network
by updating the weights and biases through the process of `back-propagation',
typically using the derivative of the loss function with respect to a weight
to update that weight \cite{kingma2015AdamMethod}.

\subsection{\label{cnn:networks}Network Structure}
\begin{figure}
\centering
\includegraphics[width=\columnwidth]{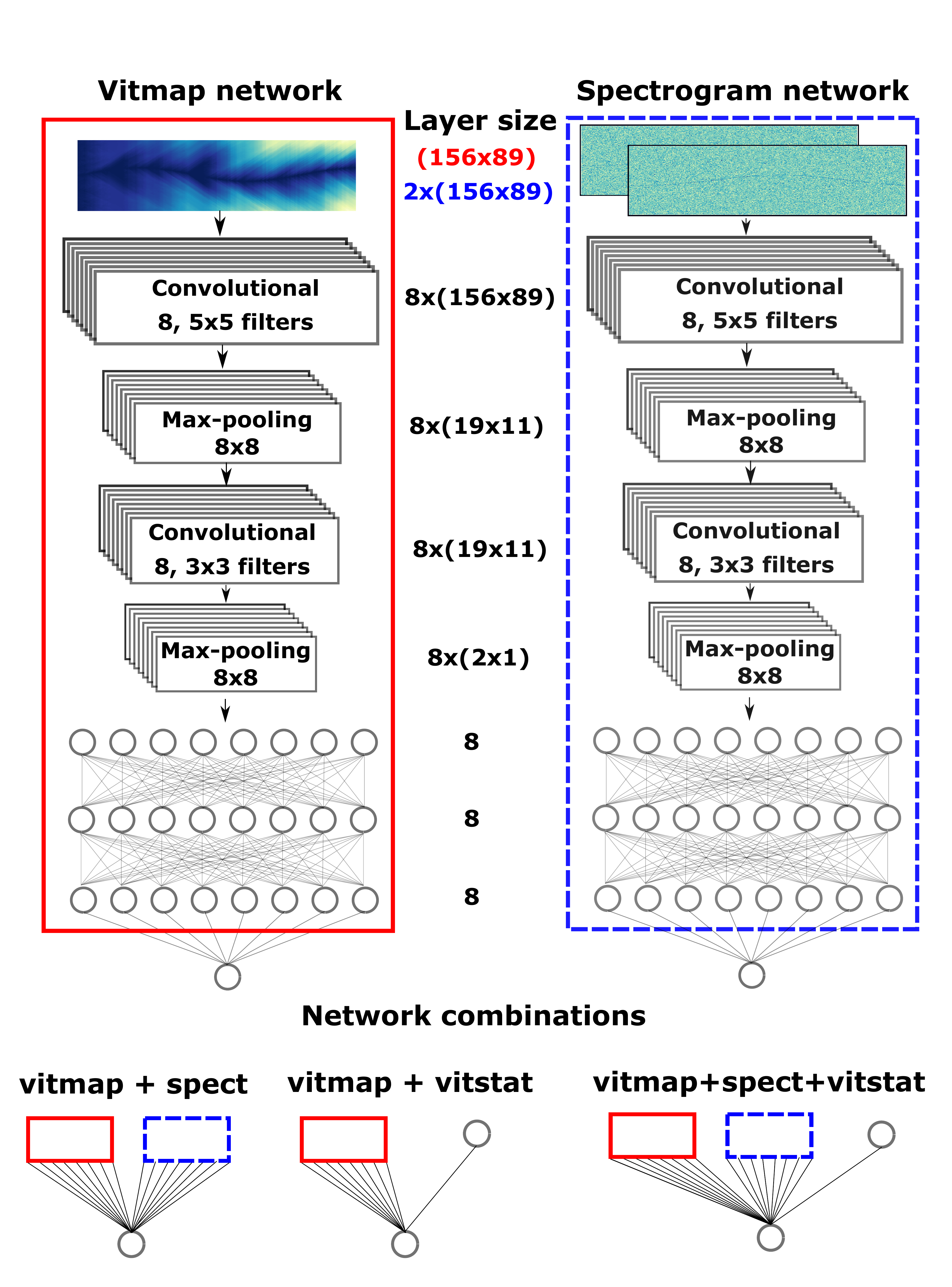}
\caption{\label{results:cnnlayout} The structure of the Viterbi map (vitmap) and spectrograms \acp{CNN} and the arrangement of the combined \acp{CNN}.
The Viterbi map and spectrogram \acp{CNN} are identical other than the input to the spectrogram \ac{CNN} is two images. They each use two
convolutional layers and 3 fully connected layers before they're output to a
single neuron which represents the probability of belonging to the signal
class. The size of the layers as the image progresses through the network is shown, where the image size is in parentheses. The Viterbi statistic network is a single neuron that transforms the
statistic into a number between 0 and 1 representing the probability of
belonging to the signal class. When multiple networks are combined,
the final output neuron and the weights connecting to the previous layer are removed, i.e. in the vitmap network the components inside the red box are used. In the vitmap + spect case, each network then has 8 output neurons which are combined to a
single neuron using 16 new weights.}
\end{figure}

In this section we describe the structure of the networks used in our
analysis. There are three possible inputs for each \ac{CNN}: a spectrogram, a
Viterbi map and the Viterbi statistic. Each of these are different
representations of the raw detector data. We proceed by training a separate
\ac{CNN} for each input separately and then a further three \acp{CNN} which
use combinations of inputs: Viterbi map + spectrogram, Viterbi map + Viterbi
statistic and Viterbi map + Viterbi statistic + spectrogram. With the
exception of the output layer, all the \ac{CNN} layers use the `leakyRELU'
activation function \cite{maas2013RectifierNonlinearities} in
Eq.~\ref{cnn:conv:equation} and \ref{machine:nn:neuron:equation}. We use a
sigmoid activation function for the output neuron so that, for a given input,
a \ac{CNN} generates an output a value between 0 and 1. The closer this
output value is to 1 the greater the probability that the input contains a
signal, so this output value can then be treated as a detection statistic.
The structure of the networks for the Viterbi map (vitmap), spectrograms and their combinations are shown in Fig.~\ref{results:cnnlayout} and
the components are described below:
\begin{description}
\item [Viterbi statistic] This is the simplest of the networks and comprises a single neuron which takes in the
    Viterbi statistic, applies a weight and bias and passes the result
    through a sigmoid function. This would give the same sensitivity as the Viterbi statistic on its own, however can now easily be combined with other networks.
\item [Viterbi map] The Viterbi map \ac{CNN} takes in a down-sampled
    Viterbi map of size (156,89) as input, described in
    Sec.~\ref{data:downsample}. As shown in Fig.~\ref{results:cnnlayout}, this \ac{CNN} consists of two convolutional
    layers and three fully-connected layers. The first layer has 8 filters
    which have a size of $5\times5$ pixels, the second layer has 8 filters
    with a size of $3\times3$ pixels. After each of these layers we use a
    max-pooling layer with a size of $8\times8$ pixels. This is then passed
    into three fully-connected layers which all have 8 neurons and used
    leakyRELU activation functions. Finally these lead to an output neuron
    which uses a sigmoid activation function.
\item [Spectrogram] The spectrogram \ac{CNN} takes down-sampled
    spectrograms of size (156,89) as inputs (see
    Sec.~\ref{data:downsample}). It has an identical structure to the
    Viterbi map \ac{CNN} but takes the spectrograms of two different
    detectors as inputs.
 \end{description}
The next three networks are constructed from combinations of these single
\acp{CNN}:
  \begin{description}
\item [Viterbi map and spectrogram] To combine the spectrogram and Viterbi
    map network we remove the final output neuron and its 8 weights from
    each of the networks and combine these to a single sigmoid neuron with
    16 new weights.
\item [Viterbi map and Viterbi statistic] In this network we combine the
    Viterbi statistic with the Viterbi map. As before, this uses the
    pre-trained Viterbi map and Viterbi statistic \acp{CNN}. Again, the
    output sigmoid neuron and corresponding weights are removed from each
    network. The 8 neurons from the Viterbi map network and the single
    neuron from the Viterbi statistic network are then combined to a single
    neuron with 9 new weights.
\item [Viterbi map, Viterbi statistic and spectrogram] This combination
    takes all component \acp{CNN} from above. As before the final sigmoid
    output and the corresponding weights from each network are removed. The
    8 neurons from the Viterbi map and spectrograms \acp{CNN} and the
    single neuron from the Viterbi statistic are then joined into a single
    output neuron with 17 new weights.

\end{description}
To combine \acp{CNN} we use `transfer
learning'~\cite{pratt1993DiscriminabilitybasedTransfer}by taking the pre-trained
weights of the networks as a starting point for further training. In our
examples we found that we could fix the weights inside the pre-trained
networks and just train the final 16 output weights from the neurons as in
Fig.~\ref{results:cnnlayout}. We chose to investigate combinations of
networks because different representations of the data should contain
slightly different information on the presence of a signal. For example, the
Viterbi statistic contains no information on the structure of the track in
the time-frequency plane, and the Viterbi maps lose some information about
multiple lines in the band. The spectrograms contain the most information but
in an unprocessed form. When used in combination, the resulting \ac{CNN}
should be able to pick the important features from each of these
representations.

%

\section{\label{data} Data generation}
%
To train the \acp{CNN} we need to generate many examples of labelled data
corresponding to the three data inputs used above, i.e., Time-frequency
spectrograms, Viterbi maps and Viterbi statistics. The training data needs to
include many examples of possible features which could appear, such as
Gaussian noise, non-Gaussian artefacts and \ac{CW} signals. Non-Gaussian
artefacts are difficult to simulate, but it is possible to use artefacts in
real data as part of the training set. Therefore, for the majority of the
analysis that follows, the time-frequency spectrograms  used to generate the
Viterbi data come from real LIGO observing runs (see Sec.~\ref{results}).

Overall, we need to consider three sets of data,, labelled `training data',
`test data' and `search data'. Training data contains a set of augmented (see
Sec.~\ref{data:augmentation}) time-frequency spectrograms containing
simulated signals and is used to train each of the networks. Test data is a
separate set of simulations which are not augmented. These are used to
generate efficiency curves and test the network. Search data does not contain
any simulated signal injections and is used to search for real signals within
the data.

%
When training and testing a network it is important that the networks are not
trained and tested on the same data. Otherwise the \acp{CNN} can learn
specific features of the training data and not the underlying distribution of
features. To avoid this, the spectrograms are split into sub-bands of width
$0.1$\,Hz. Alternating bands are designated as `odd' or `even', so that bands
starting at 100.1, 100.3\,Hz are odd and those starting at 100.2, 100.4\,Hz
are even etc. The networks can then be trained on the odd bands and tested on
the even bands and vice versa. When we search over data we will therefore
have two  trained networks, one for the even bands and one for the odd bands.

\subsection{\label{data:injections} Signal simulations}
%
To inject simulated signals into real data we first generate a set of signals
with parameters drawn randomly from prior distributions defined in
Table~\ref{data:injections:table}. The \ac{SNR} of the simulations is
uniformly distributed between 50 and 150. Where the \ac{SNR} is the coherently `recovered'
\ac{SNR} defined in Eq.~\ref{results:snr}. This is calculated for each time
segment using the definition of optimal \ac{SNR} in
\cite{prix2007SearchContinuous}, the total \ac{SNR} is then the sum of the
squares of these. The \ac{GW} amplitude $h_{0}$ is scaled based on the noise
\ac{PSD} to achieve this \ac{SNR}. The power spectrum of the signal can then
be simulated in each time segment of a time-frequency spectrogram. This is
done by assuming that the spectrogram is $\chi^2$ distributed. The the
antenna pattern functions are taken into account for the given source
parameters and detector such that the \ac{SNR} for each time segment is
calculated. This \ac{SNR} is spread over neighbouring frequency bins dependent
on its location in frequency. The power spectrum values can then be drawn
from a non-central $\chi^2$ distribution with the non centrality parameter
equal to the square of the \ac{SNR}. Each signal is simulated in two
detectors: \acp{LIGO} H1 and L1. The \acp{SNR} reported below are then the
sum of the squares of the \acp{SNR} from each detector. The simulation code used in this analysis can be found in \cite{bayley2020Soapcw}.

%
\begin{table*}
\caption{\label{data:injections:table} The upper and lower limits bounding
the random signal parameter. The parameters $\alpha,\sin{\left(\delta
\right)},f,\;\log{\left( \dot{f} \right)},\; \cos{\left(\iota \right)},\;
\phi_0,\; \psi$ were sampled uniformly between these ranges. The frequencies
$f_{\rm l}$ and $f_{\rm u}$ refer to the lower and upper frequency of the
band into which each signal is injected. Excluding the distribution of
frequencies $f$, all the injections parameters are sampled from the same
distributions as the S6 \ac{MDC}~\cite{walsh2016ComparisonMethods}.}
\bgroup
\def\arraystretch{1.5}
\centering
\begin{tabular}{c c c c c c c c r|}
\hline
\hline
 & $\alpha$ [rad]& $\sin\left(\delta \right)$ [rad] & $f$ [Hz]&
$\log_{10}\left(\dot{f} [\rm{Hz/s}]\right)$ & $\cos{\iota}$ [rad]& $\phi$ [rad]& $\psi$ [rad]\\
\hline
lower bound & $0$ & $-1$ & $f_{\rm l} + 0.25$ & $-9$ & $-1$ & $0$ & $0$ \\
\hline
upper bound & $2\pi$ & $1$ & $f_{\rm u} - 0.25$ & $-16$ & $1$ & $2\pi$ & $\pi/2$ \\
\hline
\end{tabular}
\egroup
\end{table*}

\subsection{\label{data:augmentation} Augmentation}
%
To train a neural network, many examples of data from each class are needed
to avoid over-fitting. Simply using data between 40 and 500\,Hz and splitting
the data into 0.1\,Hz wide sub-bands does not give enough data for the
networks to be trained effectively. We therefore use the technique of  data
augmentation~\cite{patrice1991TangentProp,baird1992DocumentImage} to
artificially increase the number of training examples. Augmentation is the
process of transforming existing data so that, to the network, it appears to
be `new' data. For example, by reversing a time-frequency band in time we
get a new realisation of noise in that frequency band. This gives two noise realisations for each frequency band and would double the size of the training data-set, reducing the
likelihood of over-fitting to the training data.

\begin{figure}
\centering
\includegraphics[width=\columnwidth]{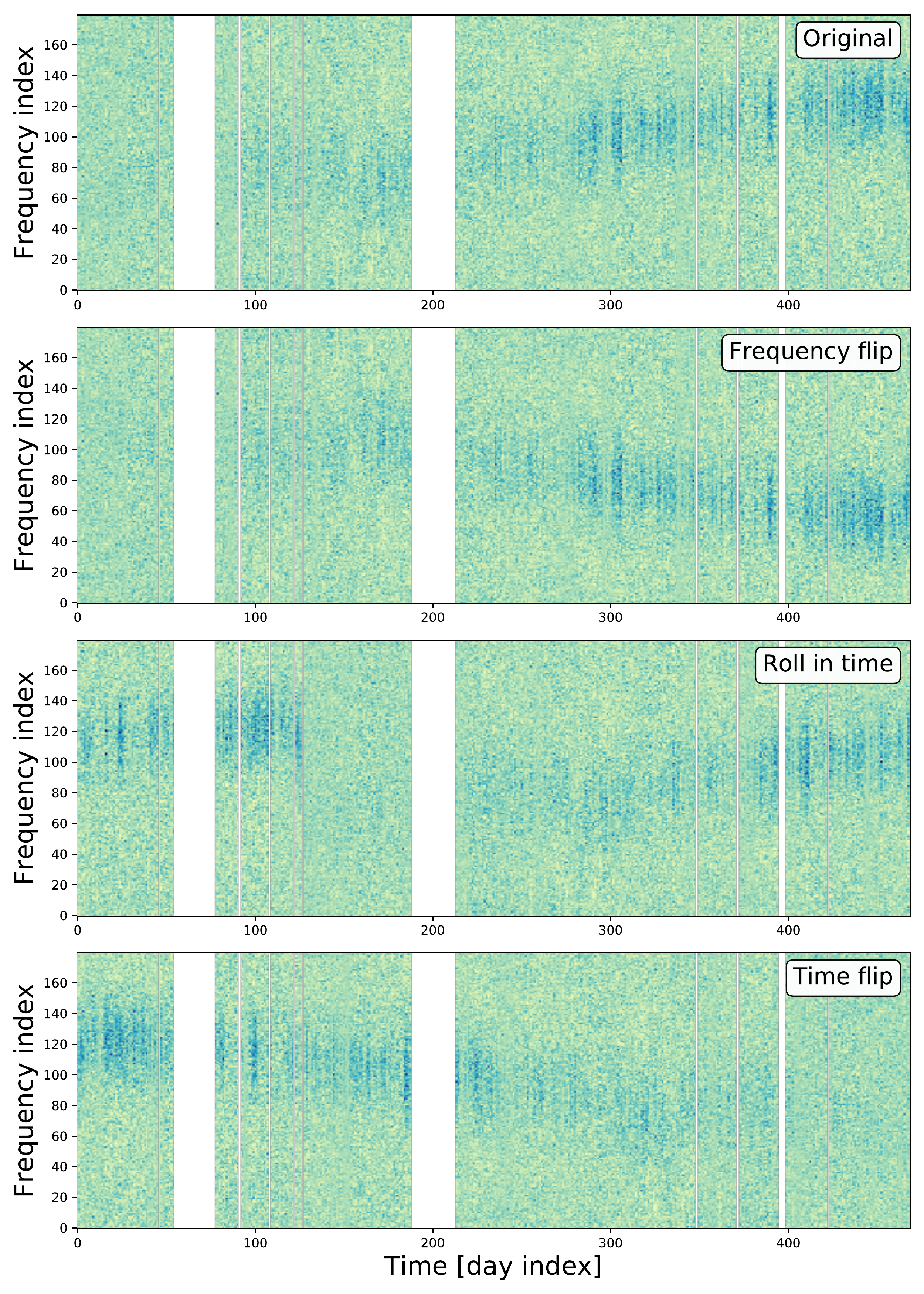}
\caption{The data is transformed by flipping the data in frequency (panel 2),
rolling the data in time by 100 bins (panel 3) and flipping the data in time
(panel 4). The original summed spectrogram is show in panel 1. Simulated
signals can then be injected using this data as noise. The plots above show a
broad wandering line to demonstrate the changes to the data when it is
augmented, however, the majority of sub-bands contain almost Gaussian
noise. }
\label{data:augmentation:examples}

\end{figure}

%
We applied augmentations to the spectrograms from each of the detectors. The
augmentations that are used on each sub-band are: reversing the data in time,
flipping the data in frequency, rolling the data in time by a small number of
segments and shifting the data in frequency by a small number of bins. As we
use real data, there are gaps in time where the detectors were not operating.
We preserve the location of these gaps when augmenting the data. When
shifting the data in frequency we shift each band up and down by 30 frequency
bins (0.016\,Hz) and up and down by 60 frequency bins (0.032\,Hz). When
rolling the data in time, we roll each sub-band by 100 time segments (100
days). Fig.~\ref{data:augmentation:examples} shows examples of the original
data, a flip in frequency, a roll in time and a flip in time. For each
frequency shift, we flip the sub-band in time and frequency and roll the
sub-band in time. This then gives us three transformations for each of the
four frequency shifts, which including the original data gives 15 augmentations of each band and therefore 15 times the
number of training examples.

\subsection{\label{data:downsample} Downsampling}
%

The raw spectrograms contain a large number of pixels and, as the
spectrograms pass through the network, there are a correspondingly large
number of computations to perform and a significant burden on memory. To
reduce their size,  the spectrograms are binned in time over one day, i.e.,
every 48 time segments, as in~\cite{bayley2019SOAPGeneralised}. As well as
reducing the size of the spectrogram, this increases the \ac{SNR} within a
given time-frequency bin assuming that the signal remains within the
frequency bin for the majority of the time segment. To reduce the size of the
data further we used the `resize' package from scikit-image
\cite{vanderwalt2014ScikitimageImage} to interpolate and resize the summed
spectrograms to 156 time segments by 89 frequency bins. This size was defined
based on the summed spectrograms of the S6 data-set. This is $1/3$ the number
of summed segments in time and $1/2$ the number in frequency. This
down-sampling is applied to the spectrograms and Viterbi maps. In \cite{bayley2019SOAPGeneralised} we
demonstrated that summing spectrograms can increase the speed and sensitivity
of our search. When down-sampling the image, we found that reducing the
amount of data had a small affect on the sensitivity of the \acp{CNN} used.

\section{\label{pipeline}Search pipeline}
%
The components described above were combined to form a single search pipeline
with a flow diagram shown in Figure~\ref{pipeline:flow} We ran this pipeline
in three modes, to train the \ac{CNN}, test the search and run a search on
real data.  The elements of the flow diagram are described below:
%
\begin{figure*}[hp]
\input{./flow.tikz}
\caption{\label{pipeline:flow} The SOAP pipeline from start to finish. There
are three main sections: Training (red), Testing (green) and Searching (grey)
for both the odd and even bands. The blue sections surrounding these indicate
that the same operation is applied to each of the training, test and search
data. The blue sections mean that the same operations are applied to all data in that section, for example, injections are made into training, test and search data for both odd and even bands in step 5.}
\end{figure*}
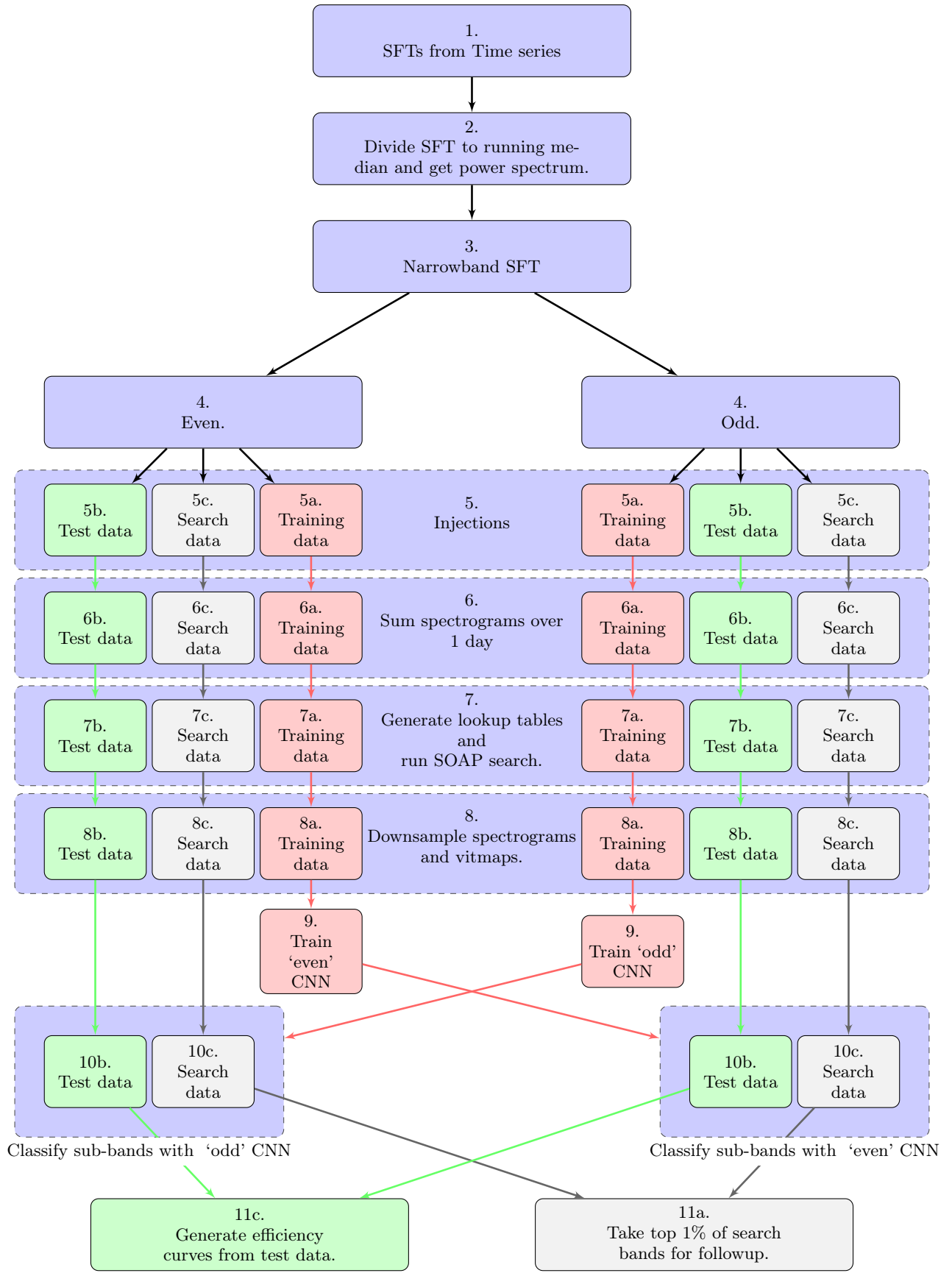

%
\begin{description}
\item[1. SFTs] These are 1\,800\,s \acp{SFT} generated from the detector
    strain time-series data. This is the standard \ac{SFT} length for a
    number of \ac{CW} searches.
\item[2. Normalising] The \acp{SFT} are then divided by their running
    median with a window width of 100 frequency bins. If we assume the
    resulting \acp{SFT} to be $\chi^2$ distributed, we can apply a
    correction factor using LALSuite code {\tt XLALSFTtoRngmed}
    \cite{ligoscientificcollaboration2018LIGOAlgorithm} such that their
    power spectrum has a mean of $\sim 1$. By then multiplying this by 2,
    the noise-like component of the spectrum is distributed as expected.
\item[3. Narrowbanding] The computational efficiency can be improved if the
    data is divided into frequency bands so the analysis can be completed
    on each band using separate CPU nodes. In this search the spectrograms
    are split into $2.1$\,Hz-wide bands every $2$\,Hz, i.e. 100.0 to
    102.1\,Hz, 102.0 to 104.1\,Hz etc. The analysis on each node will
    further split the data into 0.1\,Hz wide sub-bands. The overlap then
    allow the sub-band from 1.95-2.05 to be calculated on one node. The
    band size was chosen based on the available computational memory at the
    time.
\item[4. Band splitting] A \ac{CNN} should not be trained on the same data
    that it will be tested on, so each of the $0.1$\,Hz wide sub-bands are
    split into `odd' or `even' bands. A \ac{CNN} can then be trained on
    even bands and tested on odd bands, and vice versa.
\item[5a. Training data generation] The training data generation is
    described in Sec.~\ref{data}. Each of the $0.1$\,Hz sub-bands is
    `augmented' (Sec.~\ref{data:augmentation}). For each of the augmented
    bands, the data is duplicated and signals are injected into the copy
    with \acp{SNR} in the range 50-150 to give an example of a noise class
    member and a signal class member. There are two of these sets, one for
    `even' bands and one for `odd'.
\item[5b. Testing data generation] The signals in the testing data followin
    the parameters in Tab.~\ref{data:injections:table} are injected in to
    50\% of the $0.1$\,Hz sub-bands. These signals have an \ac{SNR} in the
    range 20-200. The \ac{SNR} range here is wider than the training set to
    show how the trained networks perform on this wider range. Again we
    have a set for `odd' and a set for `even' sub-bands.
\item[5c. Search data] This data is generated to assess the performance of
    the trained network with real signals. The sub-bands described in part
    4 are now overlapping by 0.05\,Hz. This means that if there is an
    astrophysical signal it should be fully contained within at least one
    sub-band. We do assume that the signal frequency does not change by
    more than 0.1\,Hz over a year, which is reasonable for isolated
    neutrons stars $<500$\,Hz. There are both `odd' and `even' versions of
    this search data.
\item[6. Summing spectrogram] Following the practice
    in~\cite{bayley2019SOAPGeneralised} the spectrograms are summed over
    one day, i.e., we sum 48 contiguous 30-minute time segments of the
    spectrogram to give one time segment per day. This is done separately
    for each of the six data-sets (three for `odd', three for `even').
\item[7. Generate lookup tables and run SOAP search]  When the SOAP search is run using the line-aware statistic, lookup tables which contain values of the statistic as a function of the spectrogram power in each detector \cite{bayley2019SOAPGeneralised} are used to increase the speed of the analysis. These lookup tables are generated in
    advance of the search. Once done, we run separate SOAP search for each
    of the six data-sets (three `odd', three `even') separately.
\item[8. Down-sample data] At this stage there are four saved elements for
    each of the six data-sets:  two spectrograms, the Viterbi map and the
    Viterbi statistic. The spectrograms and the Viterbi map are
    down-sampled to a size of ($156\times 89$) using interpolation from
    scikit-image's resize~\cite{vanderwalt2014ScikitimageImage}. This size
    was chosen based on the S6 \ac{MDC} data-set, where this is 1/3 the
    length in time and 1/2 the width in frequency of the summed
    spectrograms. This was chosen such that the \acp{CNN} trained
    efficiently and still achieved a reasonable sensitivity.
\item[9. Train networks] The down-sampled training data is then used to
    train the \acp{CNN}. One \ac{CNN} is trained on `odd' bands and another
    \ac{CNN} with the same structure is trained on `even' bands.
\item[10b. Run search on testing data] The trained \acp{CNN} from part 9
    are then used to classify each sub-band in the testing data with
    injections. This returns a statistic in the range $[0,1]$, where values
    closer to one imply that an astrophysical signal is likely to be
    present in the data. Here the \ac{CNN} trained on the `odd' bands is
    tested using the `even' bands and vice versa. The algorithms are run on
    this test data to asses the sensitivity of the analysis.
\item[10c. Run search on real data] The trained \acp{CNN} from part 9 are
    then used to classify each sub-band in the search data, returning a
    statistic in $[0,1]$ as in part 10b. Once again the \ac{CNN} trained on
    the `odd' bands is run on the `even' bands and vice
    versa.
\item[11a. Signal candidates] The sub-bands which return a statistic in the top
    1\% of all sub-bands can be taken as potential candidates.
    This can then be followed-up with other \ac{CW} search methods.
\item[11c. Efficiency curves] The output statistics from the test data-set
    (11b.) are examined to assess how well the network has classified
    signals as a function of the injected signal \ac{SNR}. A range of
    efficiency curves are generated, as detailed in Sec.~\ref{sensitivity}.
\end{description}

\section{\label{results}Results}
%
The networks described in Sec.~\ref{cnn:networks} were trained and tested on
four different data-sets: the S6 \ac{MDC} as
in~\cite{bayley2019SOAPGeneralised,walsh2016ComparisonMethods}, our own
injections into O2 data, Gaussian noise with the same time gaps and noise
floor as the S6 data-set, and our own injections into real S6 data. Each of
the searches uses training and testing data in the frequency range of
100-400\,Hz, except the S6 \ac{MDC} which uses data in the range 40-500\,Hz
for consistency with other searches in the challenge. 
All of the networks were trained using the Adam optimiser \cite{kingma2015AdamMethod} with a learning rate of 0.001. For each training epoch the training data was split into random batches of size 1000, where the network weights are updated after each batch. The networks were trained for 400, 200 and 4000 epochs for the vitstat, vitmap and spectrogram networks respectively.

\subsection{\label{sensitivity} Sensitivity}
%
To investigate the sensitivity of the pipeline we use two measures: the
sensitivity depth $\mathcal{D}$ \cite{prix2007SearchContinuous} and the
optimal \ac{SNR} $\rho$ \cite{behnke2015PostprocessingMethods}, both
described in \cite{bayley2019SOAPGeneralised}. The sensitivity depth is
defined as
\begin{equation}
\label{results:depth}
\mathcal{D}(f) = \frac{\sqrt{S_h(f)}}{h_0},
\end{equation}
where $S_h(f)$ is the single-sided noise \ac{PSD} and $h_0$ is the \ac{GW}
amplitude. The optimal \ac{SNR} is defined as
\begin{equation}
\label{results:snr}
\rho^2 = \sum_X 4
\Re\int^{\infty}_{0}\frac{\tilde{h}^X(f)\tilde{h}^{X*}(f)}{S^X(f)}df,
\end{equation}
where $X$ indexes over detectors and $\tilde{h}(f)$ is the Fourier transform
of the time series of the signal $h(t)$. $\rho^2$ is defined
in~\cite{prix2007SearchContinuous} for a double-sided \ac{PSD} but here we
have defined it for the more common single-sided case.

\begin{figure*}
\subfloat[]{\label{results:snr_o2}%
\includegraphics[width=\columnwidth]{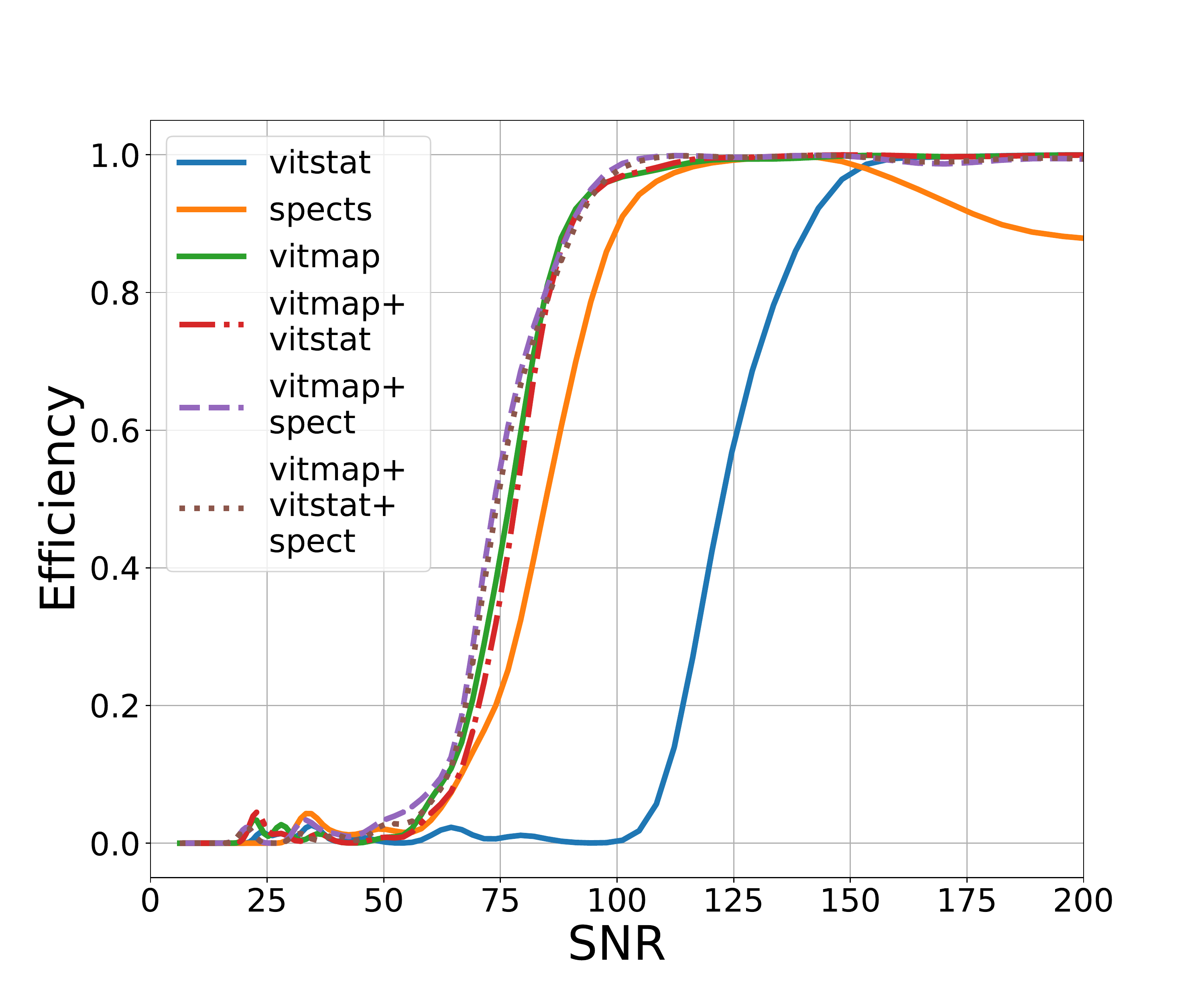}
}
\subfloat[]{\label{results:depth_o2}%
\includegraphics[width=\columnwidth]{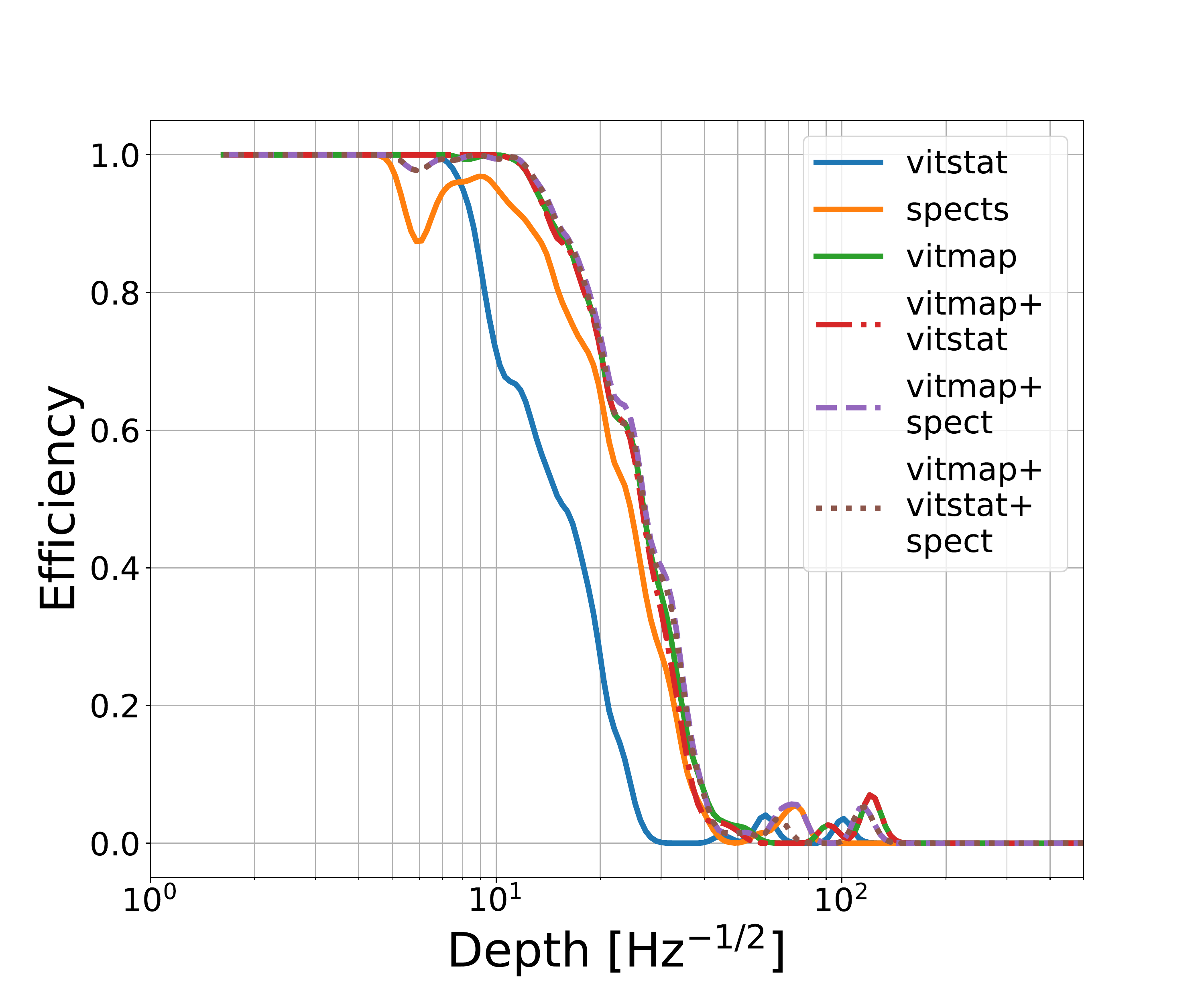}
}
\caption{\label{results:o2}  Tests of the six \acp{CNN} with the O2
data-set between 100-400 Hz. The efficiency plots above are for a 1\% false alarm probabilities.
Fig.~\ref{results:snr_o2} shows the efficiency of the search as a function of
\ac{SNR} and Fig.\ref{results:depth_o2} shows the efficiency as a function of
sensitivity depth. The efficiency here is a measure of the fraction of events
which exceed the 1\% false alarm probability for any given \ac{SNR}. These
plots both show that the sensitivity of the Viterbi statistic is
significantly below that of the different \acp{CNN}. The others group with a
similar sensitivity. }
\end{figure*}

\begin{figure*}
\subfloat[]{\label{results:snr_s6}%
\includegraphics[width=\columnwidth]{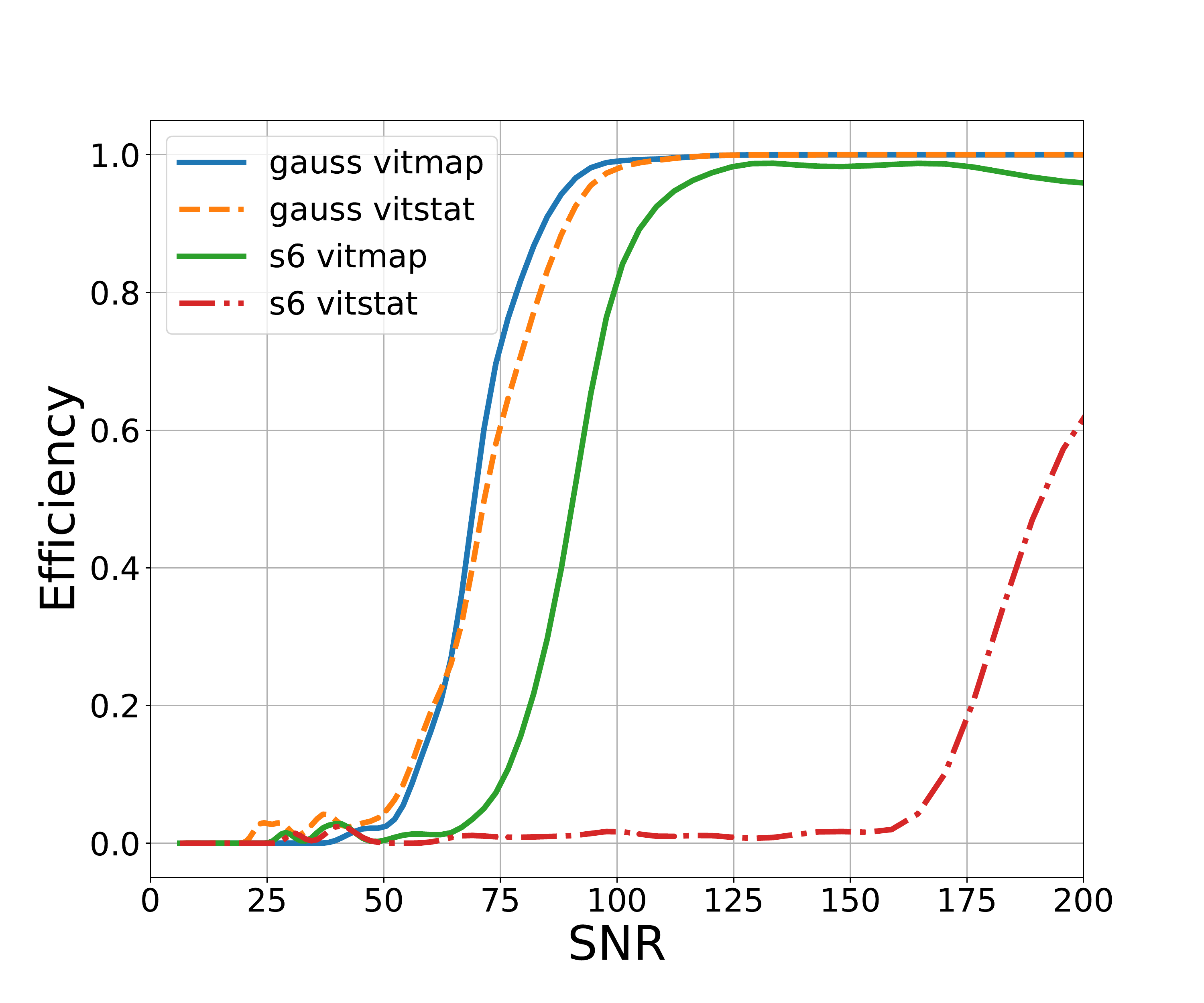}
}
\subfloat[]{\label{results:depth_s6}%
\includegraphics[width=\columnwidth]{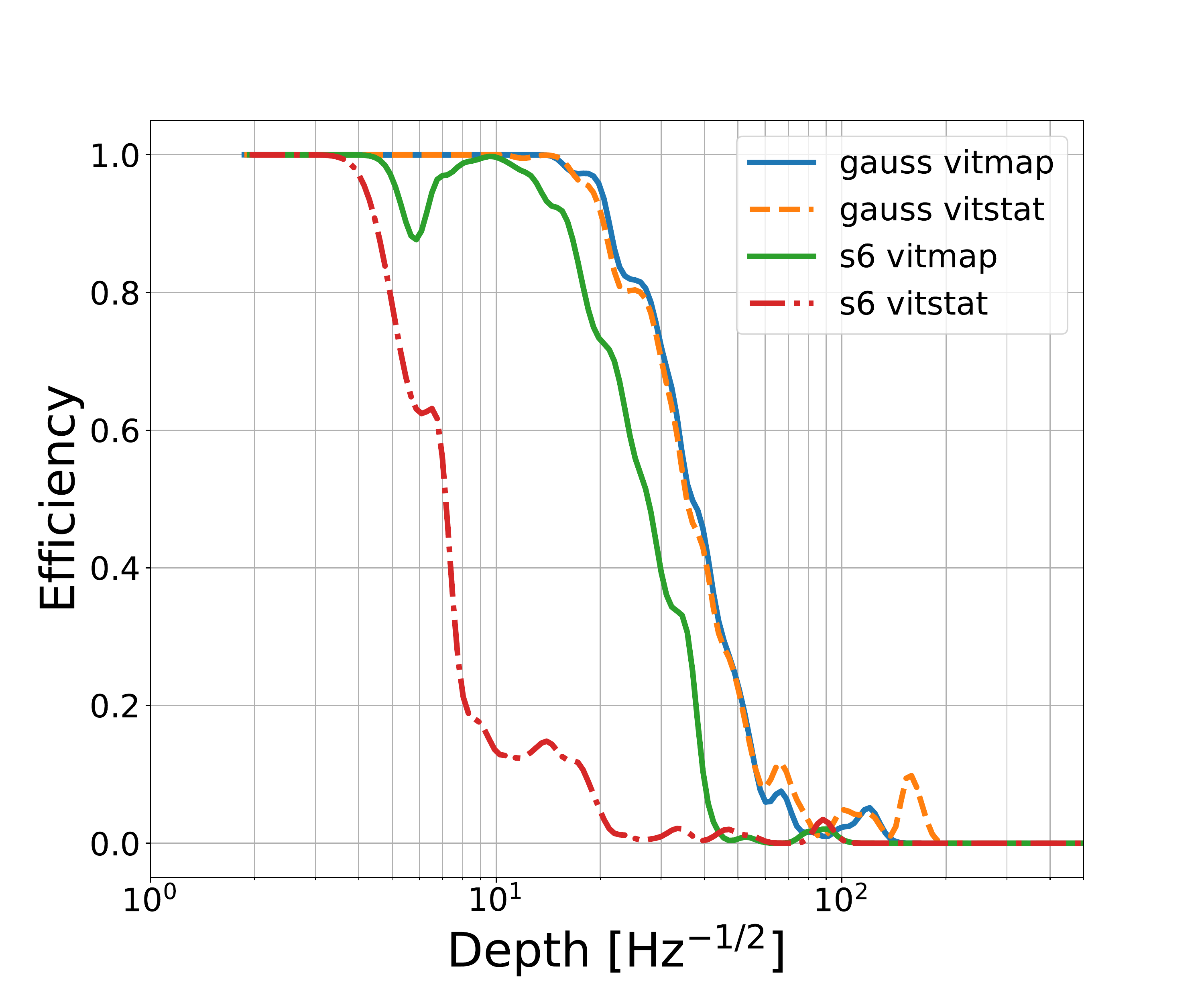}
} \caption{\label{results:s6gauss} The sensitivity of the
search for simulations in real S6 data (s6) compared to simulations in
Gaussian noise (gauss) between 100-400 Hz. Figure \ref{results:snr_s6} shows the efficiency of
the search as a function of \ac{SNR} and Fig.~\ref{results:depth_s6} shows
the efficiency as a function of sensitivity depth. The efficiency is the
fraction of events which exceed the 1\% false alarm threshold for a given
\ac{SNR} or depth. The Gaussian noise injections included the same gaps in
data as the S6 data set and the \ac{SNR} of the simulated signal in Gaussian
noise was adjusted to replicate the expected \ac{SNR} in S6 data. In the Gaussian noise
simulations the searches achieve an efficiency of $95\%$ with 1\% false alarm
at an \ac{SNR} $\sim 85$ and $\sim 90$ for the Viterbi map and Viterbi
statistic respectively. In the real S6 noise simulations the searches achieve
an efficiency of $95\%$ with 1\% false alarm at an \ac{SNR} $\sim 108$ and $
> 200$ for the Viterbi map and Viterbi statistic respectively.}

\end{figure*}

\begin{figure*}
\subfloat[]{\label{results:snr_s6mdc}%
\includegraphics[width=\columnwidth]{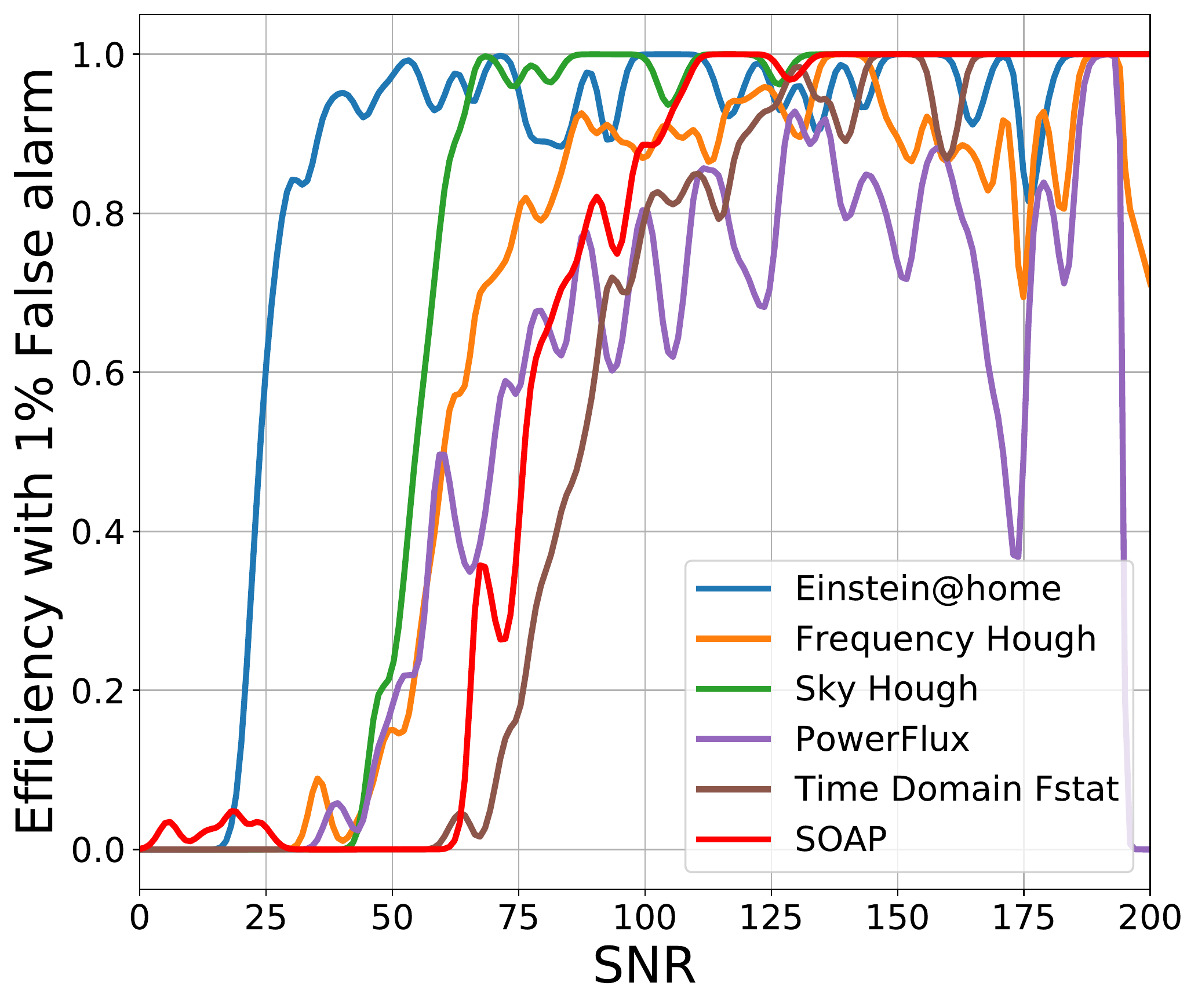}
}
\subfloat[]{\label{results:depth_s6mdc}%
\includegraphics[width=\columnwidth]{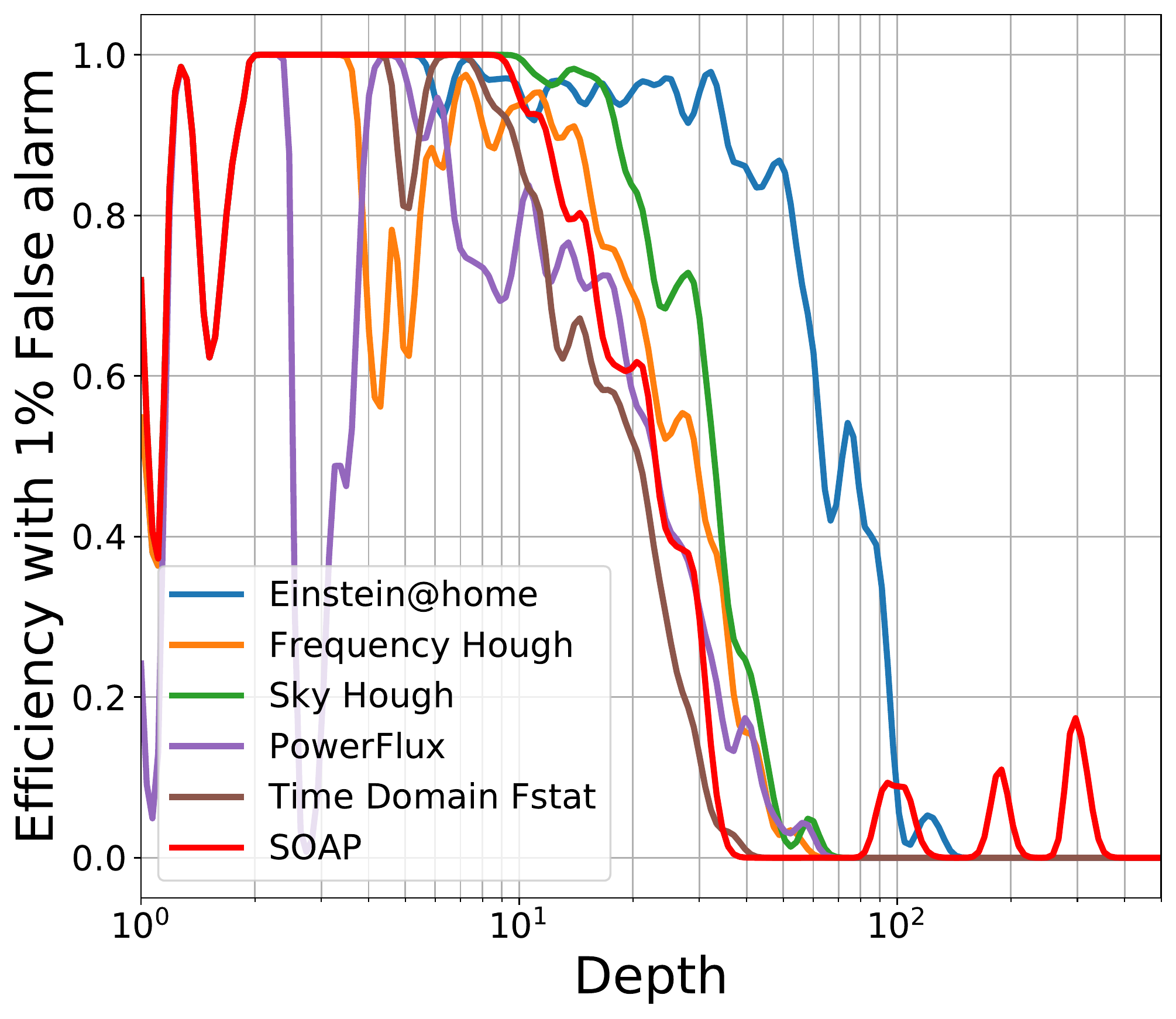}
} \caption{\label{results:s6mdc} A comparison of the SOAP + \ac{CNN} search
with other \ac{CW} searches through a standard set of injections used in the
S6 \ac{MDC} \cite{walsh2016ComparisonMethods}. We have taken the published
list of detected sources for each search \cite{walsh2016ComparisonMethods}
and replotted using the method in Sec.~\ref{sensitivity} to compare the
sensitivities to the SOAP + \ac{CNN} search. This includes results for all
source simulations between 40 and 500\,Hz. The efficiency curves are
generated with a 1\% false alarm probability. The curves are substantially more noisy than in Fig.~\ref{results:o2} or \ref{results:s6gauss} as there is a smaller number simulations in a given SNR range. }

\end{figure*}

%
The sensitivity curves shown in Fig.~\ref{results:o2},\ref{results:s6gauss}
and \ref{results:s6mdc} were generated using a $1\%$ false alarm probability, which
we use to set our detection threshold.  This threshold is the statistic value
exceeded by just $1\%$ of sub-bands that do not contain an injection. The
efficiency is defined as the fraction of events which exceed the false alarm
threshold for any given \ac{SNR}. The \ac{SNR} is sampled
uniformly between the range 20-200 as described in Sec.~\ref{pipeline}.
Instead of having multiple simulations for a discrete set of \acs{SNR} we
define a window around a point in \ac{SNR} and count the fraction of
statistics which exceed the threshold determined by the false alarm probability within that window.  We
define the window as a Gaussian with a standard deviation of 2, which is wide
enough to contain enough injections at a given \ac{SNR} to return a reliable
value. The detection efficiency $y(\rho)$ is
\begin{equation}
\label{results:gauss_smooth}
y(\rho) = \frac{\sum_i H(O_i - O^{1\%}) \mathcal{G}(\rho_i;\mu=\rho,\sigma=2)}{\sum_i
\mathcal{G}(\rho_i;\mu=\rho,\sigma=2)},
\end{equation}
where $O_i$ is the output statistic from the \ac{CNN}, $O^{1\%}$ is the
statistic value corresponding to a 1\% false alarm probability, $H$ is the Heaviside
step function which has a value of 1 for positive input arguments and 0 for
negative arguments. The \ac{SNR} of a simulation with output $O_i$ is defined
in Eq.~\ref{results:gauss_smooth} using $\rho_i$. The centre of the window in
\ac{SNR} is then $\rho$. The window is a
Gaussian with a mean of the current \ac{SNR} and a standard deviation of 2,
$\mathcal{G}(\rho_i, \mu = \rho,\sigma=2)$.
The sensitivity curves for each of the described data-sets are shown in
Figs.~\ref{results:o2},\ref{results:s6gauss} and \ref{results:s6mdc}.

%

\begin{figure}
	\includegraphics[width=\columnwidth]{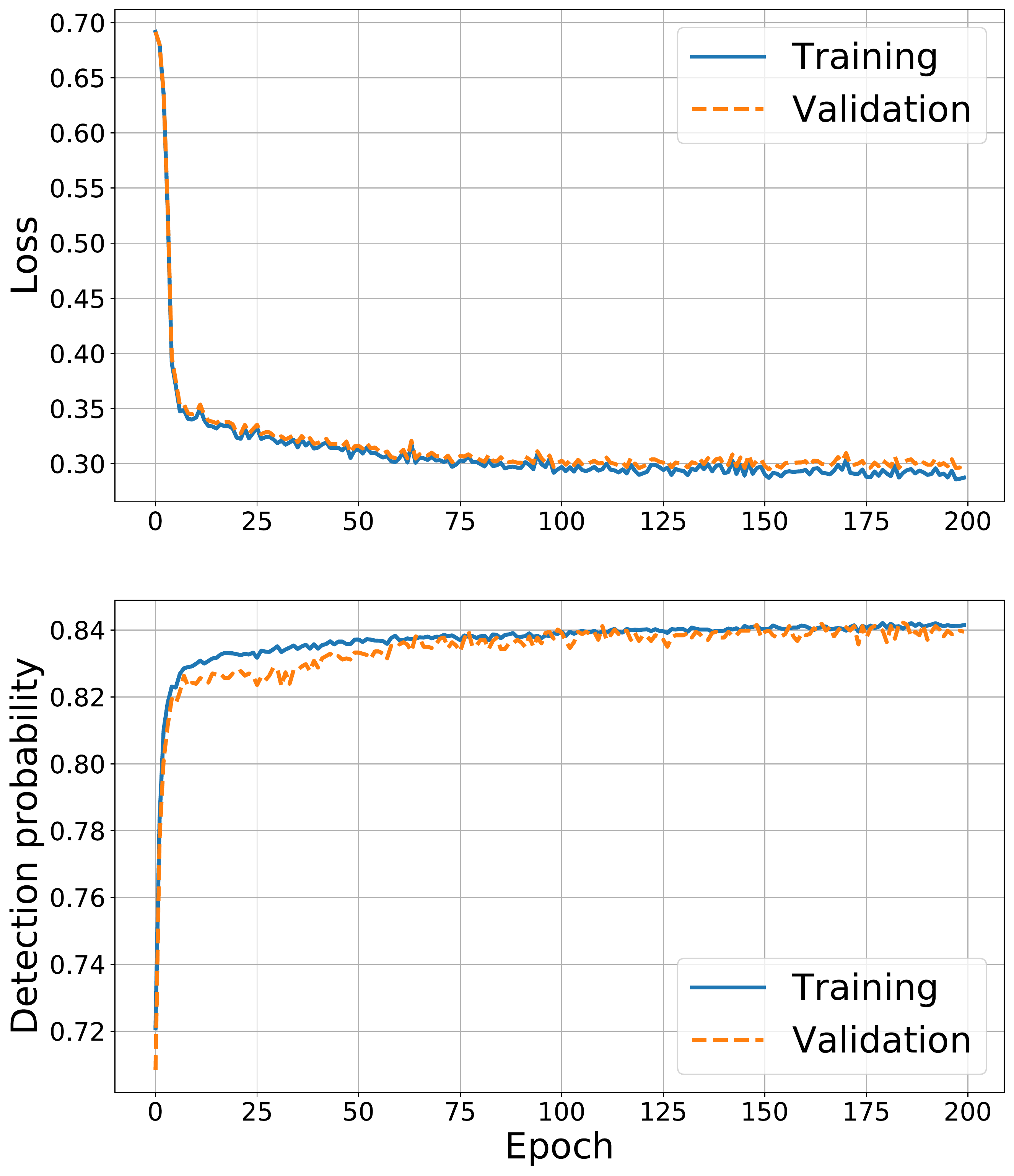}
	\caption{\label{results:loss} The loss and detection probability for training and validation sets as a function of training epoch for the vitmap network in the O2 data set. The loss decreases for both datasets with each epoch where the validation set converges after $\sim 100$ training epochs. The detection probability is calculated as the fraction of all signal simulations which exceed the 1\% false alarm value. }
\end{figure}

For the first test, injections were made into the O2 data-set as described in
Sec.~\ref{data} between 100\,Hz and 400\,Hz. Each of the six networks
described in Sec.~\ref{cnn:networks} were then trained and tested on this
data. 
Fig.~\ref{results:loss} shows an example of the training and validation loss and detection probability as a function of training epoch for the vitmap network trained on simulations in O2 data. One epoch is when the entire training set has been passed through the network. Fig.~\ref{results:loss} shows that for both the training and an independent validation set, the loss and detection probability both converge and perform similarly on each data-set, implying that the network does not over-fit.

Figure ~\ref{results:o2} shows the sensitivity curves for the tests in O2 for
both \ac{SNR} and sensitivity depth for each of the six networks. Focusing on
Fig.~\ref{results:snr_o2}, the least sensitive of the \acp{CNN} is the
Viterbi statistic (vitstat), and this is expected. We know that,  despite the
line-aware component to the Viterbi statistic calculation, it can still fail
to distinguish between some instrumental lines and astrophysical signals. The
spectrogram \ac{CNN} has an improved sensitivity over the Viterbi statistic;
this importantly does not involve the SOAP search but is run entirely on
down-sampled and summed spectrograms. This network is approaching the most
sensitive of the examples in Fig.~\ref{results:o2}. The difference in 
sensitivity between the spectrogram and the Viterbi map \ac{CNN} is most likely 
due to the Viterbi map providing a distilled representation of the spectrograms which is easier for a \ac{CNN} to interpret.
It is possible that the spectrogram \ac{CNN} could reach the same sensitivity as the Viterbi map \ac{CNN} by 
changing its structure or the data-set resolution. However, as explained in more detail in
Sec.~\ref{results:timing}, this network takes $\sim10$ times longer to train
than the Viterbi map network.

The remaining four networks contain the Viterbi map (vitmap) as one of their
inputs (or their only input) and all achieve similar sensitivities. It
appears therefore that the dominant effect on sensitivity is from the Viterbi
maps component. In the following tests our focus will be on the Viterbi map
\ac{CNN} as in all cases this is competitively the most sensitive. For the O2
data-set we show that, with a false alarm probability of 1\%, the Viterbi map
\ac{CNN} achieves a sensitivity of \ac{SNR} $~95$ and sensitivity depth of
$~12\; {\rm Hz}^{-1/2}$ with 95\% efficiency. In Fig.~\ref{results:snr_o2}
the sensitivity of the spectrogram \ac{CNN} drops after an \ac{SNR} of 150.
This is most likely due to the training set containing simulations between
and \ac{SNR} of 50 and 150, and therefore has not seen signal simulations of
higher \ac{SNR}. The dip in sensitivity in Fig.~\ref{results:depth_o2} at
lower depths is due to the same effect.
%

For the second test we simulate the S6 data-set with Gaussian noise,
retaining the same gaps in the data present in S6 but without including
instrumental artefacts such as lines. The noise floor of S6 was also
replicated by scaling the \ac{SNR} of each injection by an estimate of the
\ac{PSD} at that frequency. Figure ~\ref{results:s6gauss} shows the \ac{SNR}
and depth sensitivity curves for the Viterbi statistic and Viterbi map
\ac{CNN} for both the Gaussian noise run with S6 gaps and for injections into
the real S6 data-set. In the Gaussian noise data-set the curves for both the
Viterbi map \ac{CNN} and the Viterbi statistic, show very similar results.
This is to be expected as the main use of the \ac{CNN} was to reduce the
effect of instrumental lines, and there are none in this data set. The
advantage of using the Viterbi maps in a \ac{CNN} becomes clear when it is
tested on simulations into real S6 data with many instrumental lines. The two
curves corresponding to simulations in real S6 data in
Fig.~\ref{results:snr_s6} show the sensitivity as a function of \ac{SNR} in
these tests. It becomes clear that the Viterbi map \ac{CNN} reduces the
effect of instrumental lines and increases the searches sensitivity to
\ac{SNR}. A similar feature can be seen in Fig.\ref{results:depth_s6} where
the use of an \ac{CNN} again greatly improves sensitivity.

These tests on S6 data also show that the effect of instrumental lines was
far greater in this run than in O2. This is shown in
Fig.~\ref{results:snr_o2} where the separation between the Viterbi statistic
curves and the Viterbi map curves is significantly smaller than the S6 curves
in Fig.~\ref{results:snr_s6}. For simulations into Gaussian noise following
S6 gaps we show that with a false alarm of 1\% the Viterbi map \ac{CNN}
achieves a sensitivity of SNR $~85$ and sensitivity depth of $~20\; {\rm
Hz}^{-1/2}$ with 95\% efficiency. For injections into real S6 data the search
achieves a sensitivity of SNR $~115$ and sensitivity depth of $~11\; {\rm
Hz}^{-1/2}$ with 95\% efficiency and 1\% false alarm. We can also see from
Fig.~\ref{results:snr_s6} that the sensitivity of the vitmap \ac{CNN} in
Gaussian noise with S6 gaps is better than in real S6 data. 

%
The final test  also uses the S6 data-set, however, in this case we use the
standard set of injections used in previous \ac{CW} analysis pipeline
comparisons\ac{MDC}~\cite{walsh2016ComparisonMethods}.
Fig.~\ref{results:s6mdc} shows the resultant sensitivity curves derived from
these injections. In both Fig.~\ref{results:snr_s6mdc} and
\ref{results:depth_s6mdc} the sensitivity curves are substantially more noisy
than in Fig.~\ref{results:o2} or \ref{results:s6gauss}, mainly due to the
smaller size of the testing set. The standard set of simulations in
Fig.~\ref{results:s6mdc} contained $\sim 900$ signal simulations between 40
and 500\,Hz where the majority of these signals are distributed between an
\ac{SNR} of 0 and 150. Figures \ref{results:o2} and \ref{results:s6gauss} are
generated using $2\,300$ simulations between 40 and 500\,Hz and \acp{SNR} of
20 and 200 as described in Sec.~\ref{pipeline}. Figure
~\ref{results:depth_s6mdc} shows the direct comparison in depth of the
results in~\cite{walsh2016ComparisonMethods} with the results from the SOAP
search with the Viterbi map \ac{CNN}. This shows that we achieve a
sensitivity consistent with that of other semi-coherent searches with the
exception of the Einstein@home search~\cite{abbott2016ResultsDeepest}. Whilst
we are not the most sensitive search by this measure, the SOAP + \ac{CNN}
search offers a greatly reduced computational cost (see
Sec.~\ref{results:timing}).

This particular test was limited to signals from isolated neutron stars.
However, unlike some other semi-coherent searches, SOAP is sensitive to a
broad range of continuous-wave signals, including binary sources and sources
with limited coherence times. The inclusion of the \ac{CNN} does introduce
some dependency on the signal model, as the training set for the \ac{CNN}
contains simulations of isolated neutron stars. However, this is not a
limitation of the method: new training sets can be readily generated using a
different signal models. For tests in
the S6 \ac{MDC} we show that with a false alarm of 1\% the Viterbi map
\ac{CNN} achieves a sensitivity in SNR of $\sim 110$ and sensitivity depth of
$\sim 10\;{\rm Hz}^{-1/2}$ with 95\% efficiency.

\subsection{\label{results:timing} Computational time}
A key parameter for any \ac{CW} search is the computational time it takes to
run. Table \ref{timing:table} shows the timings for different sections of the
search using the S6 data-set. This is split into three sections: data
generation, \ac{CNN} training and \ac{CNN} testing. The majority of the
computational time taken to get from raw \acp{SFT} to results occurs is the
data generation step. The timings shown Tab.~\ref{timing:table} are for the
S6 observing run where each section is run on a single \ac{CPU} or \ac{GPU},
however, in practice the generation of the data is run on multiple \acp{CPU}
on a computing cluster. The training and testing of a \ac{CNN} is done on a
single \ac{GPU}, this substantially decreases the training time compared to a
\ac{CPU} due to the intrinsically parallel nature of neural networks.
\begin{table}
\caption{\label{timing:table} Approximate timings for training and testing
using the S6 data-set (the longest run we tested), starting from 22\,538
\acp{SFT} each of duration 1\,800\,s. The frequency range covered is
40-500\,Hz. In the training, testing and search data sections we averaged the
\acp{SFT} over one day to give 469 time segments as input to the \acp{CNN}.
The data generation times quoted are for a single \ac{CPU} however, in
reality this will be split across many separate \acp{CPU}. The training and
testing is completed on a single \ac{GPU}.}
\bgroup
\def\arraystretch{1.5}
\centering
\begin{tabular}{c c c}
 \multicolumn{3}{c}{{\bf Generating data on single CPU}}  \\
\hline
\hline
 & \multicolumn{2}{c}{Time [hrs]} \\
\hline
Narrow-banding & \multicolumn{2}{c}{$\sim 9$}   \\
\hline
Training data& \multicolumn{2}{c}{$\sim 240$} \\
\hline
Testing data& \multicolumn{2}{c}{$\sim 75$}  \\
\hline
Search data& \multicolumn{2}{c}{$\sim 40$}  \\
\hline
\hline
 \multicolumn{3}{c}{{\bf Training \acp{CNN} on single GPU} }  \\
\hline
\hline
 & Training time [hrs] & Loading time [hrs]\\
\hline
Viterbi statistic & $0.03$ & $0.2$ \\
\hline
Viterbi map & $0.8$ & $0.7$ \\
\hline
spectrogram & $9$ & $1$\\
\hline
Viterbi map \\ + Viterbi statistic& $1$ &$0.7$ \\
\hline
Viterbi map \\ + spectrogram& $1.4$ & $1.6$\\
\hline
Viterbi map \\ + Viterbi statistic \\ + spectrogram& $1.5$ & $2$ \\
\hline
\hline
 \multicolumn{3}{c}{{\bf Testing \acp{CNN} on real data on GPU}}  \\
 \hline
 \hline
 & Testing [s] & Loading [s] \\
\hline
 All \acp{CNN} & $5$ & $60-160$ \\
\hline
\hline
\end{tabular}
\egroup
\end{table}

%
Starting with raw \acp{SFT} covering the 40-500\,Hz band of the S6 data-set
(i.e., 22\,538 time segments each 1\,800\,s long, giving 828\,000 frequency
bins in total) and without any trained networks, this search would have a
total computing time of $\sim 386$ hours on a single \ac{CPU} and \ac{GPU}. However, the majority of this time is used making the
simulated data. The generation of training, testing and search data can be
easily parallelised, and in practice this is split over 200 \acp{CPU} and
takes just $\sim 2$ real-time hours ($\sim 355$ CPU hours).
After this parallelisation, if one was given S6 data without any trained networks, the
search would then take approximately $13$ hours to get an efficiency curve
and a list of candidates. In this case I assume that only the Viterbi map
network is trained and tested based on the conclusions from
Sec.~\ref{results}.

The computational cost can be reduced further if a network has been trained
on a previous observing run, negating the need for new training data and the
training itself. This reduces the total run time on S6 to $\sim 9.5$ hours
(on 200 \acp{CPU}). The reduction is significant but not drastic, as the
majority of the time is spent narrow-banding the \acp{SFT}.

To reduce the time taken to generate results at the end of an observing run, one could
narrowband the \acp{SFT} periodically as the data is taken during an
observing run. This would allow the results to be generated within $\sim 3.5$
hours of the end of the run. \acp{SFT} generated on a regular basis would
allow results to be generated during an observing run. This could be done,
for example, on a weekly basis by adding 7 days of pixels to a spectrogram,
then retraining a \ac{CNN} and generating results.

The computational cost of this search is small when compared to other
existing \ac{CW} searches. In \cite{walsh2016ComparisonMethods} the expected
computational cost for the first 4 months of O1 for each search is shown,
where the fastest search takes $0.9$ million core-hours (Hough searches) and
the slowest is $100-170$ million core-hours (Einstein@Home). The equivalent
cost of the SOAP + \ac{CNN} search is $\sim 100-200$ core-hours which is
$\sim 5-10$ thousand times faster.

\section{Summary}

%
In this paper we summarise an extension of the SOAP algorithm
\citep{bayley2019SOAPGeneralised}. The extension makes use of a \ac{CNN} to
limit the effect of instrumental lines in searches for astrophysical \ac{CW}
signals. The SOAP search has a number of outputs and in this paper we
focussed on two of these: the Viterbi statistic and the Viterbi map. The
Viterbi statistic has previously been used as a measure of whether a given
frequency band contains an astrophysical
signal~\citep{bayley2019SOAPGeneralised}. The Viterbi map is an output map
with the same shape as the input spectrogram, with a value related to the
probability that a signal passes through any particular time-frequency bin.
We use both the Viterbi map and spectrogram as input images to a \ac{CNN} and
classify each frequency band as containing an astrophysical signal or not.
This approach removes then need to manually look through frequency bands and
remove those which are contaminated with non-astrophysical (instrumental)
features.

%
In detail, we tested six separate \acp{CNN} which take in a combination of
three representations of the input data: the Viterbi statistic, the Viterbi
map and a normalised spectrogram. The objective here was to combine these
different representations of the data to increase the robustness and
sensitivity of the search. The tests found that the \ac{CNN} which uses the
Viterbi map alone as input was more sensitive than any other which used a
single data type as input. Each of the \acp{CNN} that used a combination of
input data types had a similar sensitivity to the Viterbi map \ac{CNN}.
Therefore we concluded that the Viterbi map provides the most useful summary
of information for detecting a signal. Given that the main aim of this
investigation was to reduce the effect of instrumental lines on the SOAP
search, it is unsurprising that tests with Gaussian noise data (with no such
lines) showed the \ac{CNN} search achieved a similar sensitivity to the
Viterbi statistic alone. The tests in Gaussian noise with S6 gaps showed that
at a 95 \% efficiency and a 1\% false alarm probability the Viterbi statistic and
Viterbi map achieved a sensitivity of SNR 95 and 90 respectively. When the
same test was run in real S6 data at a 95 \% efficiency and a 1\% false alarm
probability the Viterbi statistic and Viterbi map achieved corresponding
sensitivities of SNR 300 and 120 respectively. This demonstrates that the
\ac{CNN} approach adds robustness to SOAP and regains much of the sensitivity
that would otherwise be lost to the effects of instrumental lines in real
detector data.

%
These tests were repeated using a standard set of injections into S6 data to
make a direct comparison with other \ac{CW} search pipelines. At a 95\%
efficiency and a 1\% false alarm probability the Viterbi map \ac{CNN} achieved a
sensitivity of \ac{SNR} $ \sim 110$ and a sensitivity depth $\sim 10 \;
\rm{Hz}^{-1/2}$ . We have shown that the SOAP + \ac{CNN} approach can achieve
a similar sensitivity to other semi-coherent \ac{CW} search algorithms but
with a greatly reduced computational cost.

%
This search also offers significant flexibility in the type of signal that is
searched for.  In the above examples the focus is on isolated neutron stars,
largely to make a straight comparison with other \ac{CW} searches that are
tuned for these sources. However, the search framework itself is largely
model-free and non-parametric. By changing the training sets, the same
pipeline can be optimised for different signal types, and in future work we
aim to test its ability to identify other sources of \ac{GW} such as neutron
stars in binary systems. Additionally, we will apply a more advanced Bayesian
analysis to estimate basic source parameters which would then provide crucial
information for a more sensitive search by fully-coherent pipelines.

\section{Acknowledgements}
We would like to acknowledge the continuous wave working group of
\ac{LIGO}-Virgo Collaboration for their assistance during this project. We
would also like to acknowledge Sin\'ead Walsh for providing us with the data
from the S6 \ac{MDC} paper. This research is supported by the Science and
Technology Facilities Council., J.B.\, G.W.\ and C.M.\ are supported by the Science
and Technology Research Council (grant No. ST/L000946/1).
C.M.\ is also supported by the European Cooperation in Science and Technology
(COST) action CA17137. The authors are grateful for computational resources
provided by the LIGO Laboratory supported by National Science Foundation
Grants PHY-0757058 and PHY-0823459.

\bibliography{soapcnn}

\end{document}

%% file: flow.tikz
\tikzstyle{block} = [rectangle, draw, fill=blue!20, 
    text width=17em, text centered, rounded corners, minimum height=4em]
\tikzstyle{line} = [draw,line width=0.35mm, -latex']
\tikzstyle{fillnode} = [rectangle, fill=white, text centered]

\tikzstyle{blocktrain} = [rectangle, draw, fill=red!20, 
    text width=5em, text centered, rounded corners, minimum height=4em]
\tikzstyle{blocktest} = [rectangle, draw, fill=green!20, 
    text width=5em, text centered, rounded corners, minimum height=4em]
\tikzstyle{blocksearch} = [rectangle, draw, fill=black!5, 
    text width=5em, text centered, rounded corners, minimum height=4em]
\tikzstyle{blocktestbig} = [rectangle, draw, fill=green!20, 
    text width=17em, text centered, rounded corners, minimum height=4em]
\tikzstyle{blocksearchbig} = [rectangle, draw, fill=black!5, 
    text width=17em, text centered, rounded corners, minimum height=4em]
\tikzstyle{back group} = [fill=blue!20,rounded corners, draw=black!70, dashed, inner xsep=15pt, inner ysep=7pt, text centered]
\tikzstyle{back group1} = [fill=blue!20,rounded corners, draw=black!70, dashed, inner xsep=15pt, inner ysep=15pt, text centered]

\begin{tikzpicture}[node distance = 6em, auto]

    
  \node [block] (sft) {1.\\ SFTs from Time series};
  
  \node [block, below of=sft] (norm) {2. \\ Divide SFT to running median and get power spectrum.};
  \node [block, below of=norm] (narrow) {3. \\ Narrowband SFT };
  
  \node [block, below right =1.5cm and -0.9cm of narrow] (odd) {4. \\ Odd.};
  \node [block, below left =1.5cm and -0.9cm of narrow] (even) {4. \\ Even.};
  
    
  \node [blocktest,below of= odd] (testodd) {5b.\\ Test data};
  \node [blocktest, below of= testodd] (testsumodd) {6b. \\Test data};
  \node [blocktest, below of=testsumodd] (testlookupodd) {7b. \\ Test data};
  \node [blocktest, below of=testlookupodd] (testdownsampodd) {8b. \\  Test data};
  \node [below of= testdownsampodd](testblankodd) {};
    
  \node [blocktrain, left of= testodd] (trainodd) {5a.\\ Training data};
  \node [blocktrain, below of= trainodd] (trainsumodd) {6a. \\Training data};
  \node [blocktrain, below of=trainsumodd] (trainlookupodd) {7a. \\ Training data};
  \node [blocktrain, below of=trainlookupodd] (traindownsampodd) {8a. \\  Training data};
  \node [blocktrain, below of=traindownsampodd] (trainnetworkodd) {9. \\ Train `odd' \ \ac{CNN}};

  \node [blocksearch,right of=testodd] (searchodd) {5c.\\ Search data};
  \node [blocksearch, below of= searchodd] (searchsumodd) {6c. \\Search data};
  \node [blocksearch, below of=searchsumodd] (searchlookupodd) {7c. \\ Search data};
  \node [blocksearch, below of=searchlookupodd] (searchdownsampodd) {8c. \\  Search data};
  \node [below of= searchdownsampodd](searchblankodd) {};
  
  \node [blocktest, below = 1.4cm of testblankodd] (testclassifyodd) {10b.\\ Test data};
  \node [blocksearch, below = 1.4cm of searchblankodd] (searchclassifyodd) {10c.\\ Search data};

  
    \node [blocksearch,below of=even] (searcheven) {5c.\\ Search data};
  \node [blocksearch, below of= searcheven] (searchsumeven) {6c. \\Search data};
  \node [blocksearch, below of=searchsumeven] (searchlookupeven) {7c. \\ Search data};
  \node [blocksearch, below of=searchlookupeven] (searchdownsampeven) {8c. \\ Search data};
  \node [below of= searchdownsampeven](searchblankeven) {};
  
  \node [blocktest,left of= searcheven] (testeven) {5b.\\Test data};
  \node [blocktest, below of= testeven] (testsumeven) {6b. \\Test data};
  \node [blocktest, below of= testsumeven] (testlookupeven) {7b. \\Test data};
  \node [blocktest, below of=testlookupeven] (testdownsampeven) {8b. \\ Test data};
  \node [below of= testdownsampeven](testblankeven) {};
  
  \node [blocktrain, right of= searcheven] (traineven) {5a.\\ Training data};
  \node [blocktrain, below of= traineven] (trainsumeven) {6a. Training data};
  \node [blocktrain, below of=trainsumeven] (trainlookupeven) {7a. \\Training data};
  \node [blocktrain, below of=trainlookupeven] (traindownsampeven) {8a. \\ Training data};
  \node [blocktrain, below of=traindownsampeven] (trainnetworkeven) {9. \\ Train `even'  \ \ac{CNN}};

  \node [blocktest, below = 1.4cm of testblankeven] (testclassifyeven) {10b.\\ Test data};
  \node [blocksearch, below = 1.4cm of searchblankeven] (searchclassifyeven) {10c.\\ Search data};
  
  
\begin{scope}[on background layer]
   
    \node (bkgen) [back group] [fit=(trainodd) (testodd) (searchodd) (traineven) (testeven) (searcheven) ] {5.\\Injections};
    
    \node (bksum) [back group] [fit=(trainsumodd) (testsumodd) (searchsumodd) (trainsumeven) (testsumeven) (searchsumeven)] {6.\\Sum spectrograms over \\1 day};
    
    \node (bksoap) [back group] [fit=(trainlookupodd) (testlookupodd) (searchlookupodd) (trainlookupeven) (testlookupeven) (searchlookupeven)] {7.\\Generate lookup tables \\and\\ run SOAP search.};
    
    \node (bkdown) [back group] [fit=(traindownsampodd) (testdownsampodd) (searchdownsampodd) (traindownsampeven) (testdownsampeven) (searchdownsampeven)] {8.\\Downsample spectrograms\\ and vitmaps.};
    
    \node (bkclassodd) [back group1] [fit=(testclassifyodd) (searchclassifyodd)] {};
     
    \node (bkclasseven) [back group1] [fit=(testclassifyeven) (searchclassifyeven)] {};

 \end{scope}
 
  
  \node [blocktestbig, below right =1.1cm and -3.5cm of bkclasseven] (output) {11c.\\ Generate efficiency curves from test data.};
  \node [blocksearchbig, below left= 1.1cm and -3.5cm of bkclassodd] (outputsearch) {11a.\\ Take top 1\% of search bands for followup.};
  
  \path [line] (sft) -- (norm);
  \path [line] (norm) -- (narrow);
  
  
   \path [line] (narrow) -- (even);
  
  \path [line] (even) -- (testeven);
  \path [line] (even) -- (traineven);
  \path [line] (even) -- (searcheven);
  
  \path [line,red!60] (traineven) -- (trainsumeven.north);
  \path [line,green!60] (testeven.south) -- (testeven.south|-testsumeven.north);
  \path [line,black!60] (searcheven.south) -- (searcheven.south|-searchsumeven.north);
  
  \path [line,green!60] (testeven.south|-testsumeven.south) -- (testeven.south|-testlookupeven.north);
  \path [line,red!60] (traineven.south|-trainsumeven.south) -- (traineven.south|-trainlookupeven.north);
  \path [line,black!60] (searcheven.south|-searchsumeven.south) -- (searcheven.south|-searchlookupeven.north);
  
  \path [line,green!60] (testeven.south|-testlookupeven.south) -- (testeven.south|-testdownsampeven.north);
  \path [line,red!60] (traineven.south|-trainlookupeven.south) -- (traineven.south|-traindownsampeven.north);
  \path [line,black!60] (searcheven.south|-searchlookupeven.south) -- (searcheven.south|-searchdownsampeven.north);
  
  \path [line,red!60] (traindownsampeven) -- (trainnetworkeven);
  \path [line,green!60] (testdownsampeven) -- (testclassifyeven);
  \path [line,black!60] (searchdownsampeven) -- (searchclassifyeven);
  
  
  \path [line,red!60] (trainnetworkeven) -- (bkclassodd);
  
  
   \path [line] (narrow) -- (odd);
  
  \path [line] (odd) -- (testodd);
  \path [line] (odd) -- (trainodd);
  \path [line] (odd) -- (searchodd);
  
  \path [line,red!60] (trainodd) -- (trainsumodd.north);
  \path [line,green!60] (testodd.south) -- (testodd.south|-testsumodd.north);
  \path [line,black!60] (searchodd.south) -- (searchodd.south|-searchsumodd.north);
  
  \path [line,green!60] (testodd.south|-testsumodd.south) -- (testodd.south|-testlookupodd.north);
  \path [line,red!60] (trainodd.south|-trainsumodd.south) -- (trainodd.south|-trainlookupodd.north);
  \path [line,black!60] (searchodd.south|-searchsumodd.south) -- (searchodd.south|-searchlookupodd.north);
  
  \path [line,green!60] (testodd.south|-testlookupodd.south) -- (testodd.south|-testdownsampodd.north);
  \path [line,red!60] (trainodd.south|-trainlookupodd.south) -- (trainodd.south|-traindownsampodd.north);
  \path [line,black!60] (searchodd.south|-searchlookupodd.south) -- (searchodd.south|-searchdownsampodd.north);
  
  \path [line,red!60] (traindownsampodd) -- (trainnetworkodd);
  \path [line,green!60] (testdownsampodd) -- (testclassifyodd);
  \path [line,black!60] (searchdownsampodd) -- (searchclassifyodd);
  
  \%path [line,green!60] (testodd.south|-downsampodd.south) -- (testodd.south|-classifyodd.north);
  
  \path [line,red!60] (trainnetworkodd) -- (bkclasseven);
  
  
  \path [line,green!60] (testclassifyeven) -- (output);
  \path [line,black!60] (searchclassifyeven) -- (outputsearch);
  
  \path [line,green!60] (testclassifyodd) -- (output);
  \path [line,black!60] (searchclassifyodd) -- (outputsearch);
  
  
   \node[fillnode,below] at (bkclasseven.south) {Classify sub-bands with \ `odd' \ac{CNN}};
   
   \node[fillnode,below] at (bkclassodd.south) {Classify sub-bands with \ `even' \ac{CNN}};

\end{tikzpicture}